\documentclass[%
 reprint,
superscriptaddress,
nofootinbib,
 amsmath,
 amssymb,
 aps,
 longbibliography,
floatfix,
]{revtex4-1}

\usepackage{graphicx}%
\usepackage{dcolumn}%
\usepackage{bm}%

\usepackage[a4paper,lmargin={2.2cm},rmargin={2.2cm},tmargin={2.2cm},bmargin={2.2cm}]{geometry}

\usepackage{natbib}
\usepackage{amsfonts}
\usepackage{amsmath}
\usepackage{amssymb}
\usepackage{graphicx} %
\usepackage{caption}
\usepackage{subcaption} %
\usepackage{booktabs} %
\usepackage{tikz}
\usepackage{lipsum}
\usepackage{xargs} %
\usepackage[
    separate-uncertainty=true, 
    round-mode=uncertainty,
    round-precision=1,
]{siunitx} %
\usepackage[export]{adjustbox}  %
\usepackage{placeins}
\usepackage{hyperref}
\usepackage{subfiles}
\usepackage[normalem]{ulem}
\usepackage{ragged2e}
\usepackage[colorinlistoftodos,prependcaption,textsize=tiny]{todonotes}

\definecolor{linkcolor}{HTML}{0b5394}

\hypersetup{
    colorlinks=true,
    linkcolor=linkcolor,     %
    citecolor=linkcolor,    %
    urlcolor=linkcolor       %
}
\newcommand{\etarel}{$\eta^{\text{rel}}$}

\newcommand{\phirel}{$\phi^{\text{rel}}$}

\newcommand{\pt}{$p_\text{T}$}

\newcommand{\jetclass}{\textsc{JetClass}}
\newcommand{\omnijetalpha}{\textsc{OmniJet}-$\alpha$}

\newcommandx{\commentjosch}[1][]{\textcolor{blue}{JB: #1}}
\newcommandx{\commentgregor}[1][]{\textcolor{violet}{GK: #1}}

\newcommand{\qcd}{$q/g$}

\newcommand{\thad}{$t\to bqq'$}
\newcommand{\tlep}{$t\to b\ell\nu$}
\newcommand{\hbb}{$H\to b\bar{b}$}
\newcommand{\hcc}{$H\to c\bar{c}$}
\newcommand{\hgg}{$H\to gg$}
\newcommand{\hqqqq}{$H\to 4q$}
\newcommand{\hlnuqq}{$H\to \ell\nu qq'$}

\begin{document}

\title{\omnijetalpha{}: The first cross-task foundation model for particle physics}

\author{Joschka Birk}
\email{joschka.birk@uni-hamburg.de}
\affiliation{
	Institute for Experimental Physics, Universität Hamburg \\
	Luruper Chaussee 149, 22761 Hamburg, Germany
}

\author{Anna Hallin}
\email{anna.hallin@uni-hamburg.de}
\affiliation{
	Institute for Experimental Physics, Universität Hamburg \\
	Luruper Chaussee 149, 22761 Hamburg, Germany
}

\author{Gregor Kasieczka}%
\affiliation{
	Institute for Experimental Physics, Universität Hamburg \\
	Luruper Chaussee 149, 22761 Hamburg, Germany
}

\begin{abstract}
        \vspace{0.3cm}
	Foundation models are multi-dataset and multi-task machine learning methods
	that once pre-trained can be fine-tuned for a large variety of downstream
	applications. The successful development of such general-purpose models for
	physics data would be a major breakthrough as they could improve the
	achievable physics performance while at the same time drastically reduce the required amount of training time and data.
	We report significant progress on this challenge on several fronts. First, a
	comprehensive set of evaluation methods is introduced to judge the quality of an encoding
	from physics data into a representation suitable for the autoregressive generation of particle jets
        with transformer architectures
	(the common backbone of foundation models).         
        These measures motivate the
	choice of a higher-fidelity tokenization compared to previous works.     
	Finally, we demonstrate transfer learning between an unsupervised problem
	(jet generation) and a classic supervised task (jet tagging) with our new \omnijetalpha\ model.
	This is the first successful transfer between two different and
	actively studied classes of tasks and constitutes a major step in the building
	of foundation models for particle physics.
        \vspace{1.0cm}
\end{abstract}

\maketitle

\begin{figure*}
	\centering
	\includegraphics[width=0.95\textwidth]{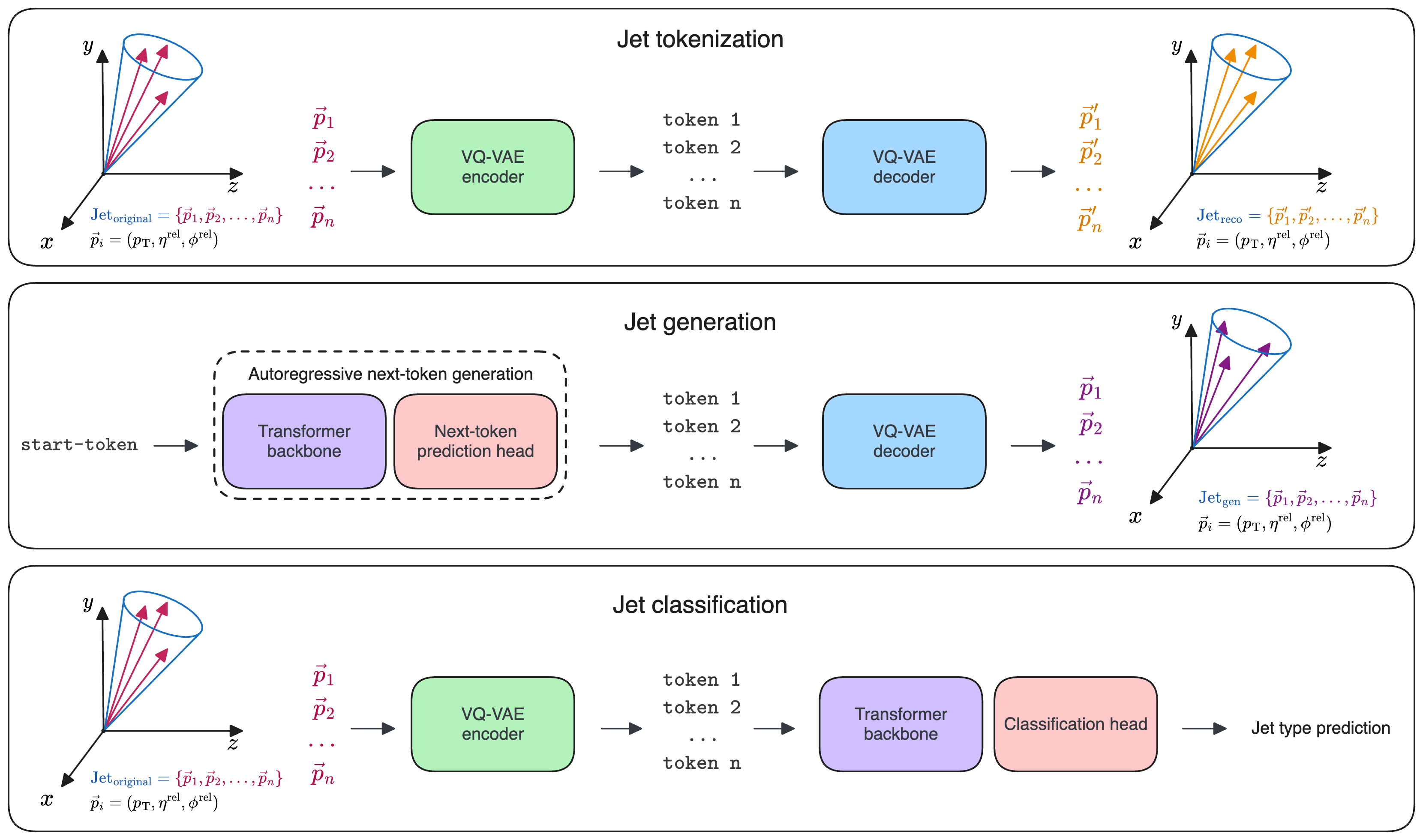}
	\caption{
		Schematics of the different steps (tokenization, generation, classification) in the \omnijetalpha\ model.
	}
	\label{fig:task_overview}
\end{figure*}

\section{Introduction}
Foundation models are a new class of machine learning models which are trained
on broad datasets and problems and are able to generalize to a variety of downstream
tasks and datasets~\cite{bommasani2022opportunities}.
Large-language models (LLMs) such as 
BERT~\cite{devlin2019bert}, BART~\cite{lewis2019bart}, GPT-3~\cite{brown2020language}, and LLaMA~\cite{touvron2023llama} 
are examples of
foundation models for text data while DALL-E~2~\cite{ramesh2022hierarchical}
is an example of an image-based model.

The benefits of a foundation model for particle physics data would be significant:
While machine learning models developed so far typically outperform classical
approaches (and often by a large
margin)~\cite{Kasieczka_2019,Karagiorgi:2022qnh}, available statistics for
training these models
is a constant issue~\cite{Macaluso_2018,qu2024particle}.
It becomes even more extreme in the case of searches for rare processes, where
only a small fraction of simulated events might pass pre-selection criteria.
Foundation models on the other hand are pre-trained and need fewer examples to
be fine-tuned for a specific task~\cite{Vigl:2024lat}.

Beyond improving the physics performance by e.g. increasing selection accuracy of classification tasks,
foundation models also address the other major issue currently facing particle
physics: limited computing resources in the face of an ever increasing amount of
data~\cite{HEPSoftwareFoundation:2017ggl,LHCC_HL:2022}.
This problem has already spawned the development of e.g. increasingly high
fidelity models for the simulation of calorimeters, as well as techniques for
the speeding up of other Monte Carlo
simulations~\cite{CaloGan1,gettinghigh,Buhmann:2023kdg,	Adelmann:2022ozp,Butter:2022rso,Hashemi:2023rgo,EventGen1,de_Oliveira_2017_learning_particle_physics_by_example}.
By allowing the re-use of models across datasets and tasks, foundation models
will play an important role in reducing this computational burden.
Note that this effect will be further compounded by the potential for
optimization in computing operations from one model used across multiple tasks.

Finally, using closely related architectures across different tasks inside one
experiment, across experimental collaborations, and in the exchange with the
theory community will make results easier to
re-interpret~\cite{araz2024les,bieringer2024classifier}.

These potential benefits have inspired research into proto-foundation models suitable
for particle physics.
For example, \cite{Dillon:2021gag,Favaro:2023xdl,Dillon:2023zac,Park:2022zov,Dillon:2022tmm} investigated
how known physical symmetries could be used to learn powerful embeddings of jet
data, \cite{Benato:2021olt} showed the versatility of graph-based message
passing networks for datasets from different domains of physics, \cite{Liu:2023lnn,Salamani:2023ttx}
demonstrated conditioning generative models on the geometry of the detector to
allow the simultaneous simulation of multiple detectors with one architecture,
\cite{Dolan:2021pml,Beauchesne:2023vie} used meta-learning for mass-decorrelation and weak-supervision,
and \cite{qu2024particle} achieved state-of-the-art performance on the top
tagging landscape dataset \cite{kasieczka_2019_2603256} by pre-training on a
different dataset~\cite{JetClass} and transferring the results.

Due to their flexibility demonstrated across language and other domains,
transformers \cite{Vaswani:2017lxt} are currently the most suitable candidate
architecture for building foundation models for applications in particle physics.
Taking inspiration from LLMs where sentences are generated autoregressively,
recent efforts have demonstrated success with
autoregressive generation of particle
physics data, for example using a
transformer to generate \thad\ and \qcd\ jets \cite{Finke:2023veq}
or to generate $Z$+jets events \cite{Butter:2023fov}. 
In \cite{heinrich2024masked} it was demonstrated how a transformer
backbone can be pre-trained using a BERT-like scheme where the model is trained to predict masked out jet constituents, resulting in an improvement of the performance when fine-tuning the backbone (with a new classification head) for jet tagging, especially at small training dataset sizes.
Furthermore, \cite{Huang:2024voo} showed
how a tokenized detector representation can be used in a BERT-like model for track reconstruction, 
and \cite{Hashemi:2023ruu} used tokenization of detector images together with a transformer to capture the context within a collider event for subsequent generation\footnote{In addition, shortly after this paper was out, \cite{Harris:2024sra} demonstrated contrastive learning with re-simulated events as a suitable pre-training task.}.

In this work, we will explore whether an autoregressive Generative Pretrained Transformer (GPT) model can be used as
a foundation model for jet physics. However, the standard GPT constructions are
not built to deal with continuous input data, but rather tokenized data.
As point clouds are the most versatile representation of physics data
\cite{Komiske_2019,Buhmann_2023,Kasieczka_2019, Buhmann:2023bwk,Buhmann:2023kdg}
and can incorporate both event level information, jet
substructure, and even low-level detector signals, finding a suitable input
transformation for point clouds to tokens is the most pressing problem. Various
tokenization strategies have been explored, for example using a simple mapping
based on binning the input space in \cite{Finke:2023veq}, a Gaussian mixture
model in \cite{Butter:2023fov}, and using an additional conditional embedding
network in \cite{heinrich2024masked}.

Here, we follow the conditional tokenization strategy from
\cite{oord2018neural,bao2022beit,heinrich2024masked}, but first take a step
back to verify the quality and trade-offs involved in building these tokens.
This will allow us to formulate quality measures to choose a suitable tokenization model,
leading to an increase in codebook size from 512 tokens
in~\cite{heinrich2024masked} to 8192 tokens.

Using this representation, we will first demonstrate training a generative model
for jets as tokens in an unsupervised way for the \jetclass\ \cite{JetClass} dataset.

Finally, this allows us to test whether the information encoded in a model
that was trained to generate jets can also be transferred to the task of classifying
them. Observing such a transfer ability across different classes of tasks --- as
opposed to transfer between different classification or generation problems ---
would be a crucial ingredient to building foundation models for physics data, and has not yet been achieved.
A graphical representation of this approach is provided in \autoref{fig:task_overview}. As this is the first prototype of a model to tackle all tasks with jets in particle physics, it is named \omnijetalpha.

The rest of the paper is organized as follows: \autoref{sec:methods}
introduces the data as well as the tokenization approach, the generative
architecture, and the transfer learning strategy. Next,
\autoref{sec:results} shows the results of the tokenization study, the
generative performance, as well as tests of the transfer learning capabilities
of the model. Finally, \autoref{sec:conclusion} summarizes the results and
provides a brief outlook.

\begin{figure*}[t]
	\centering
	\includegraphics[scale=0.25]{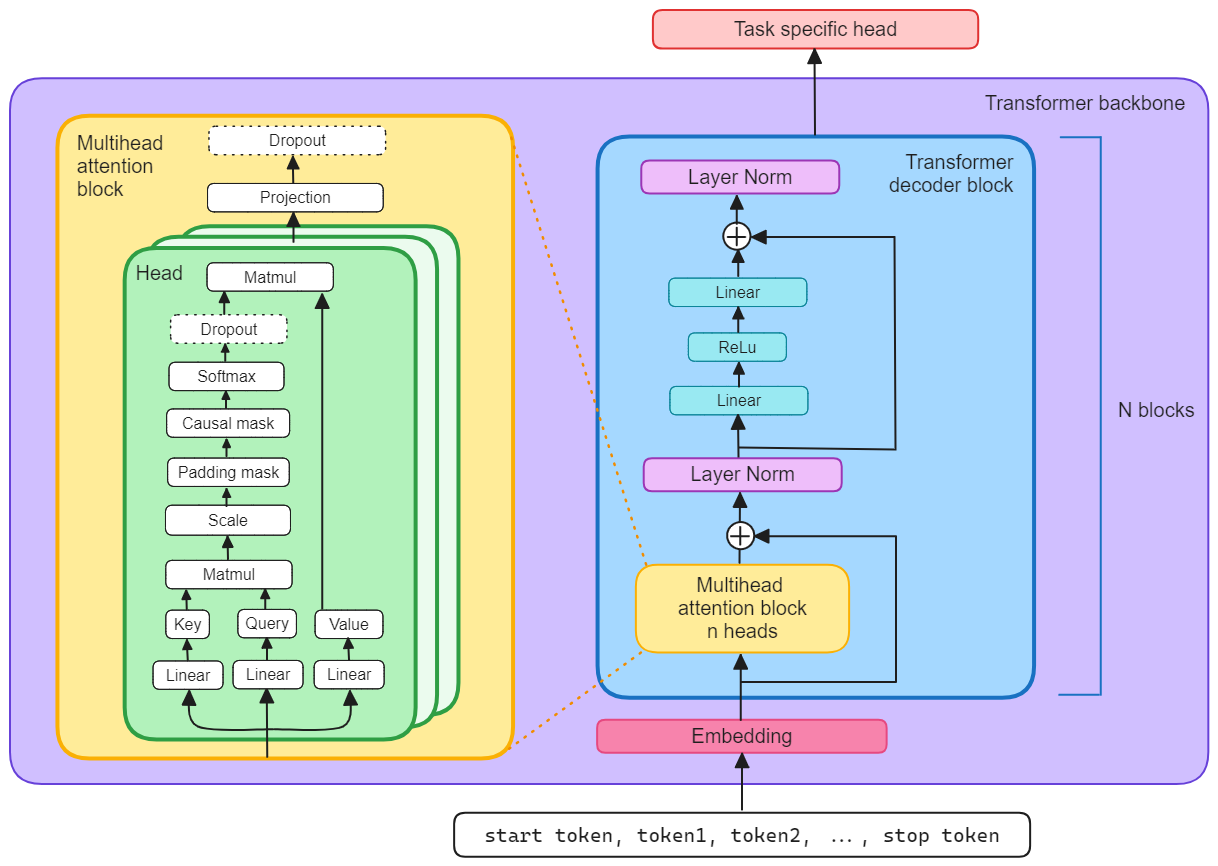}
	\caption{
		Architecture of the transformer backbone component of \omnijetalpha. The data that has been encoded by the VQ-VAE is fed through an embedding layer, before it reaches the main part of the model which is based on the transformer decoder. The output of the transformer decoder blocks is passed to a task specific head, for either generation or classification tasks. Note that during inference of the generative model, the model does not receive complete token sequences, but only the start token. The model will then autoregressively generate the rest of the sequence, updating its input as it progresses, as described in the text.
	}
	\label{fig:GPT-architecture}
\end{figure*}
\section{Methods and Dataset}
\label{sec:methods}

\subsection{Dataset}

\label{sec:data}

All studies are performed using the  \jetclass\ dataset~\cite{JetClass},
originally introduced in~\cite{qu2024particle}.
It contains both jet-level and constituent-level features
for ten different types of jets initiated by gluons and quarks ($q/g$),
top quarks ($t$, subdivided by their decay mode into \thad\ and  \tlep) , as well
as $W$, $Z$, and $H$ (\hbb, \hcc, \hgg, \hqqqq, and \hlnuqq) bosons.

Events are simulated using \textsc{MadGraph5\_aMC@NLO}~\cite{madgraph} with
parton showering and hadronization done by \textsc{Pythia}~\cite{pythia8}.
A simplified detector simulation implemented in \textsc{Delphes}~\cite{delphes}
using the CMS detector~\cite{CMS_experiment} card is performed.
Constituents are clustered into jets using the anti-$k_\mathrm{T}$
algorithm~\cite{antikt} with a distance parameter of $R=0.8$.

Jets are selected if they have a
transverse momentum of
\mbox{$ \SI{500}{GeV} < p_\mathrm{T}^\mathrm{jet} < \SI{1000}{GeV}$} and a pseudorapidity of
\mbox{$|\eta^\mathrm{jet}| < 2$}.
Additionally, truth-level matching is performed for all classes except \qcd\ and
only jets that contain all the decay products of the boson or top quark are
included.
The resulting dataset contains 100M jets for training, 5M jets for validation, and
20M jets for testing, with equal numbers of jets for each class.
This split into training, validation, and test sets corresponds to the default
split used in the \jetclass\ dataset, in order to allow for a straightforward 
comparison with previous and future works.

In this work, only the kinematic information per particle  (\pt, $\phi$, $\eta$)
is used while the particle mass $m$ is approximated as zero.
Next, the azimuth angle $\phi$ and the pseudorapidity $\eta$ are pre-processed
to be relative to the jet axis\footnote{
	The difference in $\phi$ is signed and rectified to $-\pi$
	through $\pi$. We handle those calculations using the
	\texttt{scikit-hep/vector}~\cite{schreiner_2023_7671687}
	and \texttt{scikit-hep/awkward}~\cite{pivarski_2024_10498548} packages.
}:
\begin{align}
	\eta^\mathrm{rel} & = \eta^\mathrm{particle} - \eta^\mathrm{jet}     \\
	\phi^\mathrm{rel} & = \phi^\mathrm{particle} - \phi^\mathrm{jet} \,.
\end{align}

Finally, we apply the cuts \mbox{$|\eta^\mathrm{rel}| < 0.8$} and  
\mbox{$|\phi^\mathrm{rel}| < 0.8$} to remove a very small fraction of low-energy
constituents at the periphery and use up to 128 particles per jet. 
The jet constituents are ordered by \pt{} in descending order.

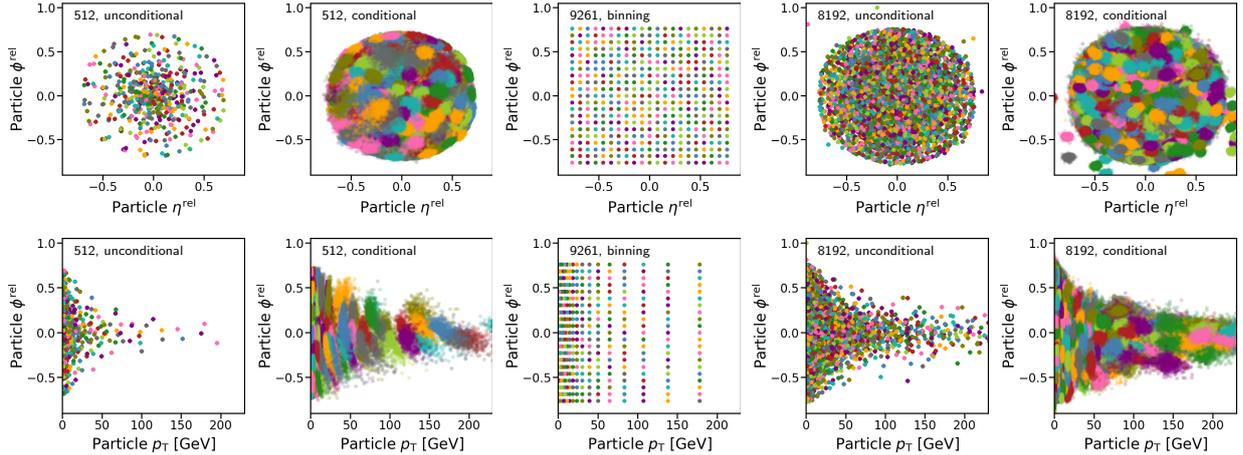
\begin{figure*}[t]
	\centering
	\begin{subfigure}[b]{0.19\linewidth}
		\begin{tikzpicture}
			\node at (0,0) {
				\includegraphics[height=3.1cm]{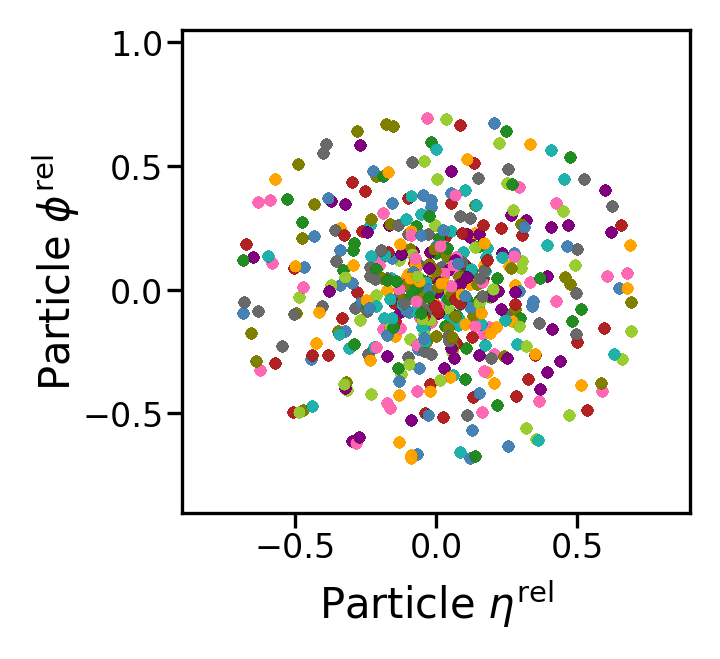}
			};
			\node[anchor=west] at (-0.8, 1.25) {\sffamily \tiny 512, unconditional};
		\end{tikzpicture} \\
		\vspace*{-0.7em}
	\end{subfigure}
	\begin{subfigure}[b]{0.19\linewidth}
		\begin{tikzpicture}
			\node at (0,0) {
				\includegraphics[height=3.1cm]{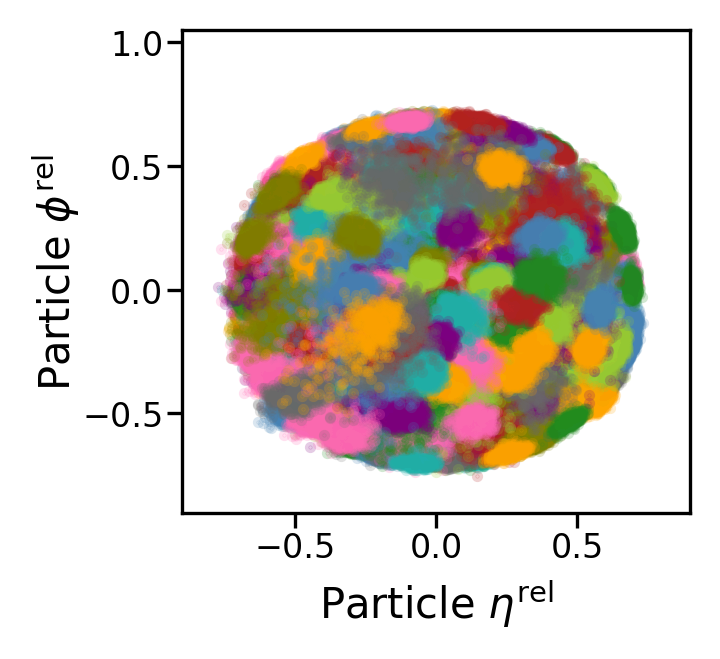}
			};
			\node[anchor=west] at (-0.8, 1.25) {\sffamily \tiny 512, conditional};
		\end{tikzpicture} \\
		\vspace*{-0.7em}
	\end{subfigure}
	\begin{subfigure}[b]{0.19\linewidth}
		\begin{tikzpicture}
			\node at (0,0) {
				\includegraphics[height=3.1cm]{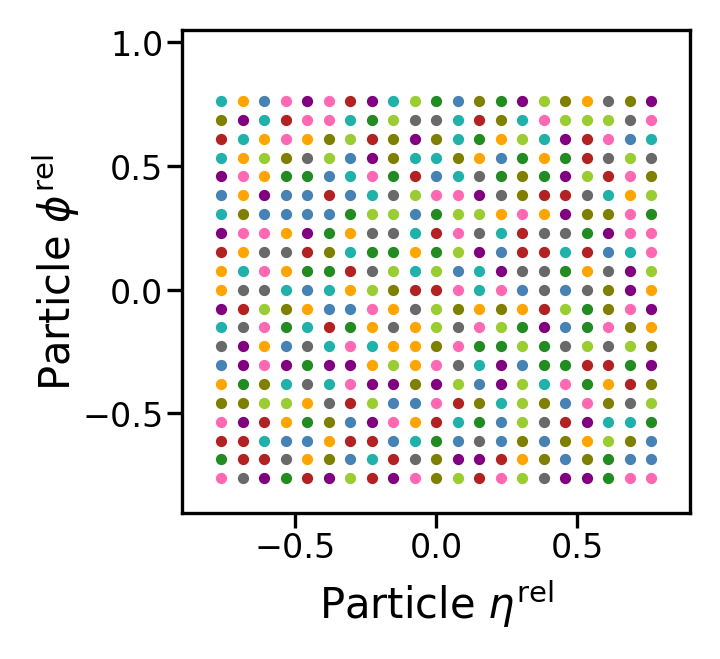}
			};
			\node[anchor=west] at (-0.8, 1.25) {\sffamily \tiny 9261, binning};
		\end{tikzpicture} \\
		\vspace*{-0.7em}
	\end{subfigure}
	\begin{subfigure}[b]{0.19\linewidth}
		\begin{tikzpicture}
			\node at (0,0) {
				\includegraphics[height=3.1cm]{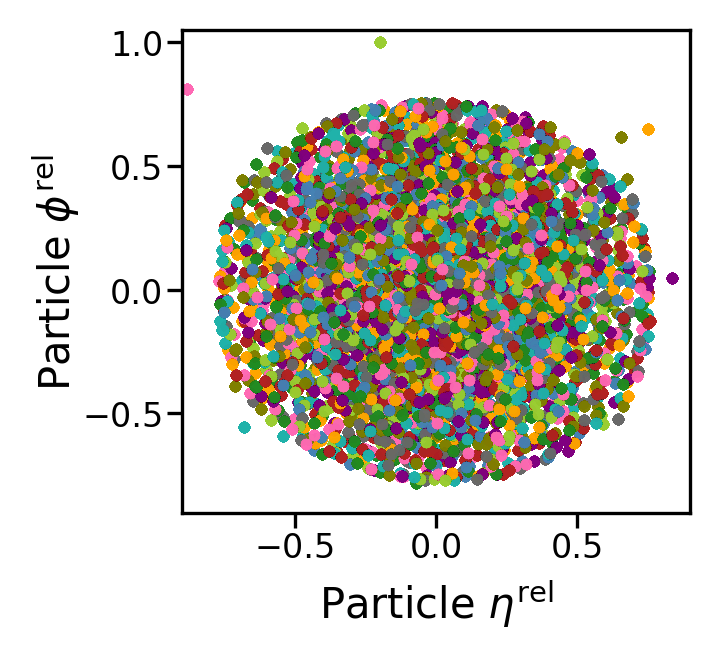}
			};
			\node[anchor=west] at (-0.8, 1.25) {\sffamily \tiny 8192, unconditional};
		\end{tikzpicture} \\
		\vspace*{-0.7em}
	\end{subfigure}
	\begin{subfigure}[b]{0.19\linewidth}
		\begin{tikzpicture}
			\node at (0,0) {
				\includegraphics[height=3.1cm]{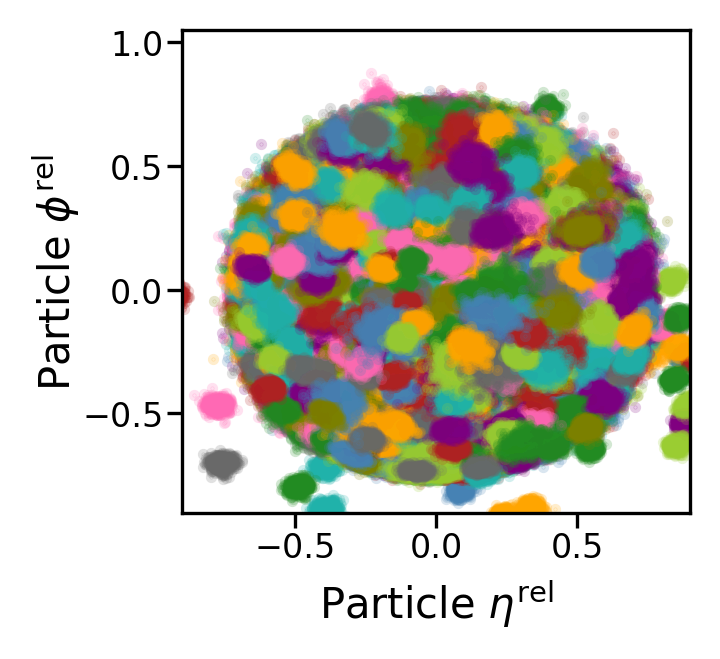} 
			};
			\node[anchor=west] at (-0.8, 1.25) {\sffamily \tiny 8192, conditional};
		\end{tikzpicture} \\
		\vspace*{-0.7em}
	\end{subfigure}
	\begin{subfigure}[b]{0.19\linewidth}
		\begin{tikzpicture}
			\node at (0,0) {
				\includegraphics[height=3.1cm]{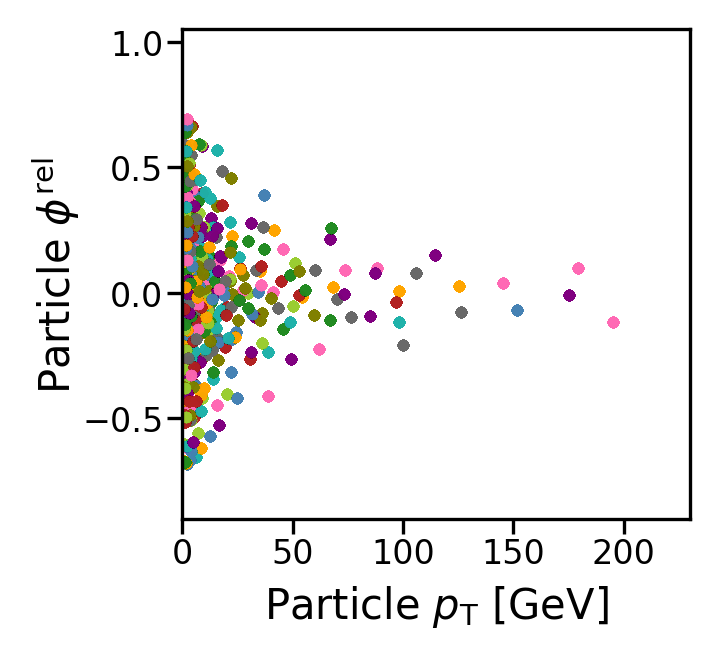}
			};
			\node[anchor=west] at (-0.8, 1.25) {\sffamily \tiny 512, unconditional};
		\end{tikzpicture} \\
		\vspace*{-0.7em}
	\end{subfigure}
	\begin{subfigure}[b]{0.19\linewidth}
		\begin{tikzpicture}
			\node at (0,0) {
				\includegraphics[height=3.1cm]{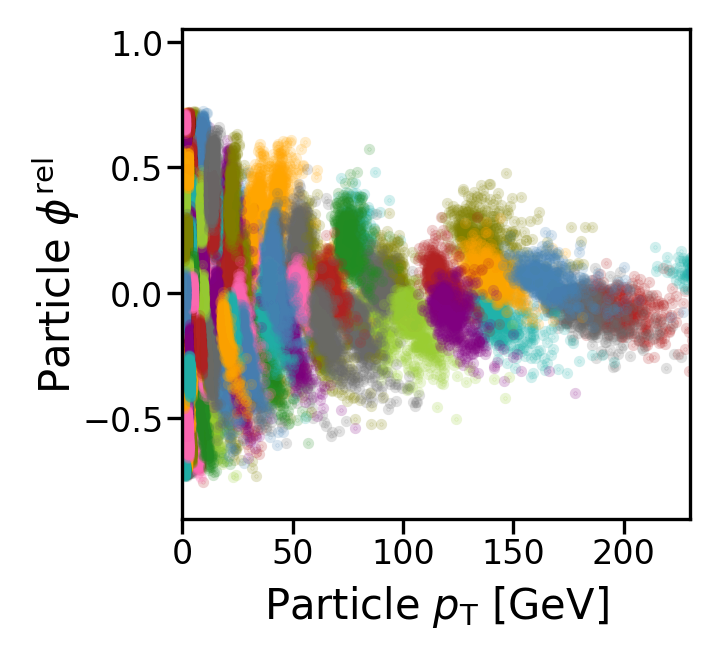} 
			};
			\node[anchor=west] at (-0.8, 1.25) {\sffamily \tiny 512, conditional};
		\end{tikzpicture} \\
		\vspace*{-0.7em}
	\end{subfigure}
	\begin{subfigure}[b]{0.19\linewidth}
		\begin{tikzpicture}
			\node at (0,0) {
				\includegraphics[height=3.1cm]{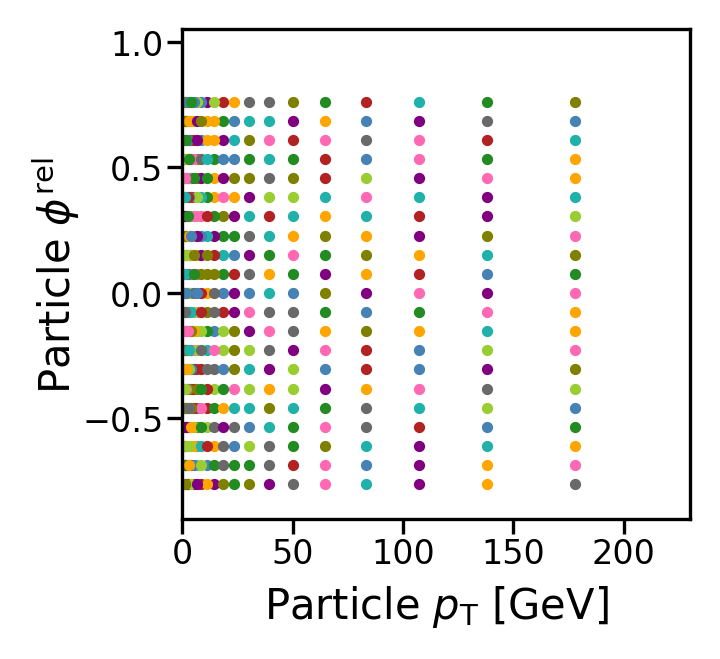}
			};
			\node[anchor=west] at (-0.8, 1.25) {\sffamily \tiny 9261, binning};
		\end{tikzpicture} \\
		\vspace*{-0.7em}
	\end{subfigure}
	\begin{subfigure}[b]{0.19\linewidth}
		\begin{tikzpicture}
			\node at (0,0) {
				\includegraphics[height=3.1cm]{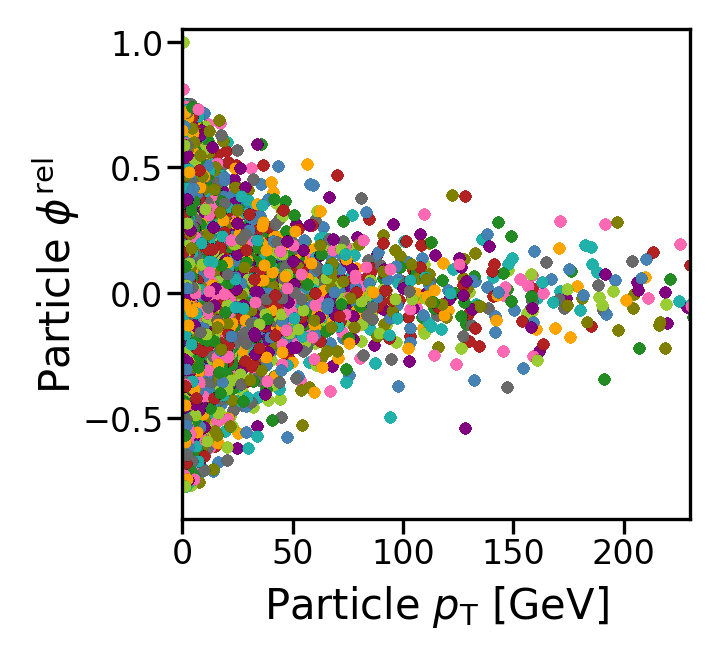} 
			};
			\node[anchor=west] at (-0.8, 1.25) {\sffamily \tiny 8192, unconditional};
		\end{tikzpicture} \\
		\vspace*{-0.7em}
	\end{subfigure}
	\begin{subfigure}[b]{0.19\linewidth}
		\begin{tikzpicture}
			\node at (0,0) {
				\includegraphics[height=3.1cm]{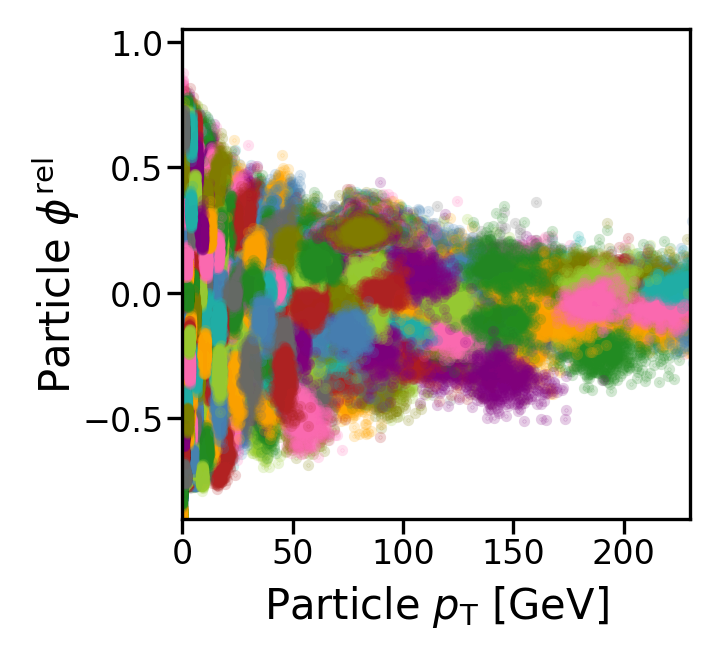}
			};
			\node[anchor=west] at (-0.8, 1.25) {\sffamily \tiny 8192, conditional};
		\end{tikzpicture} \\
		\vspace*{-0.7em}
	\end{subfigure}

	\caption{
		Visualization of the reconstructed tokens in physical space (i.e. \pt,
		\etarel, \phirel) for different tokenization approaches and codebook sizes.
		Each figure label indicates the codebook size and the tokenization approach.
		The unconditional tokenization, as well as the binning approach only have
		one reconstruction for each token, independent of the other tokens in the jet.
		To visualize the reconstruction spread of the conditional tokens,
		each token is reconstructed 500 times, each time conditioned on
		50 randomly selected tokens from the codebook. Each colored blob corresponds
		to the reconstructions obtained for one token.
	}

	\label{fig:token_visualization}
\end{figure*}

\subsection{Jet constituent token creation}

\label{sec:tokens}

We explore three kinds of tokenization approaches:
binned, conditional, and unconditional tokenization.
In the binned approach \cite{Finke:2023veq}, the space of input features is subdivided using a
regular grid in all dimensions (e.g. a 21x21x21 grid in three dimensions) and
the cells in this grid are enumerated, resulting in one token per cell.

In the unconditional approach, each constituent is tokenized individually using a
non-linear mapping, whereas in the conditional approach constituents are
encoded and decoded conditioned on each other. We use a
Vector Quantized Variational AutoEncoder 
(VQ-VAE)~\cite{oord2018neural,bao2022beit,heinrich2024masked,huh2023straightening} 
to create a discrete set of jet constituent tokens both for conditional and
unconditional tokenization.

The input features for the VQ-VAE are the \etarel, \phirel\ and \pt\ values
of the jet constituents.
For the conditional tokenization, we use a transformer for both the encoder
and the decoder of the VQ-VAE, whereas a simple multi-layer perceptron (MLP) is
used for the unconditional tokenization. Details about the different VQ-VAE
models used in our studies, as well as details about the preprocessing
of the input features can be found in \autoref{subsec:vqvae_details_appendix}.

\subsection{Transformer backbone}

\label{sec:model}

The core of \omnijetalpha{}\ is a transformer backbone based on the GPT transformer decoder model first introduced in~\cite{Radford2018ImprovingLU}. However, since jet
constituents are permutation invariant, we do not employ the positional encoding usually used in LLMs. As input, the transformer backbone receives the generated tokens from the VQ-VAE, complemented with a start and stop token. A jet with $n$ constituents is then represented as
\begin{equation}
	\left(\mathtt{start\_token}, x_1, ..., x_{n-1},x_n,\mathtt{stop\_token}\right)
\end{equation}
where $x_i$ are the tokens. 

The transformer backbone itself consists of an embedding layer followed by a series of GPT blocks. Each GPT block contains a multihead attention block, followed by a residual addition, layer norm~\cite{ba2016layer}, two linear layers with a ReLU in between, another residual addition and a final layer norm. Since this is an autoregressive model, a causal mask is passed together with the input data to the multihead attention block to prevent the model from seeing future tokens. The architecture is illustrated in \autoref{fig:GPT-architecture}.

The output from the transformer backbone is passed to a task specific head, either for generation or classification. The generative head is a single linear layer, while the classification head consists of a linear layer followed by ReLU, a sum over the constituent dimension, and a last linear layer with softmax activation function. The model is trained with $n=8$ heads in the multi-head attention block and $N=3$ GPT
blocks. No dropout is used.

Once the generative model, i.e. the transformer backbone together with the generative head, has been trained on the tokenized data, it can generate new data autoregressively. The model has learned the probability distribution for a token $x_j$, given a sequence of tokens:
\begin{equation}
	p\left(x_j|x_{j-1},...,x_1,\mathtt{start\_token}\right).
\end{equation}
The model is provided with the start token, and then samples this distribution to sequentially generate new tokens. 
Generation is repeated until either the stop token is generated or the maximum sequence length (which is set to be equal to 128) is reached. The generated token sequences are then fed to the VQ-VAE decoder, which maps them into physical space for further evaluation.

The classification task can be performed either from scratch, using randomly initialized weights for both the transformer backbone and the classification head, or by fine-tuning the generative model. In the fine-tuning case, the initial weights of the transformer backbone are loaded for from the generative model. 

\section{Results}
\label{sec:results}

\subsection{Token quality}

\begin{figure*}
	\centering
	\includegraphics[width=0.99\linewidth]{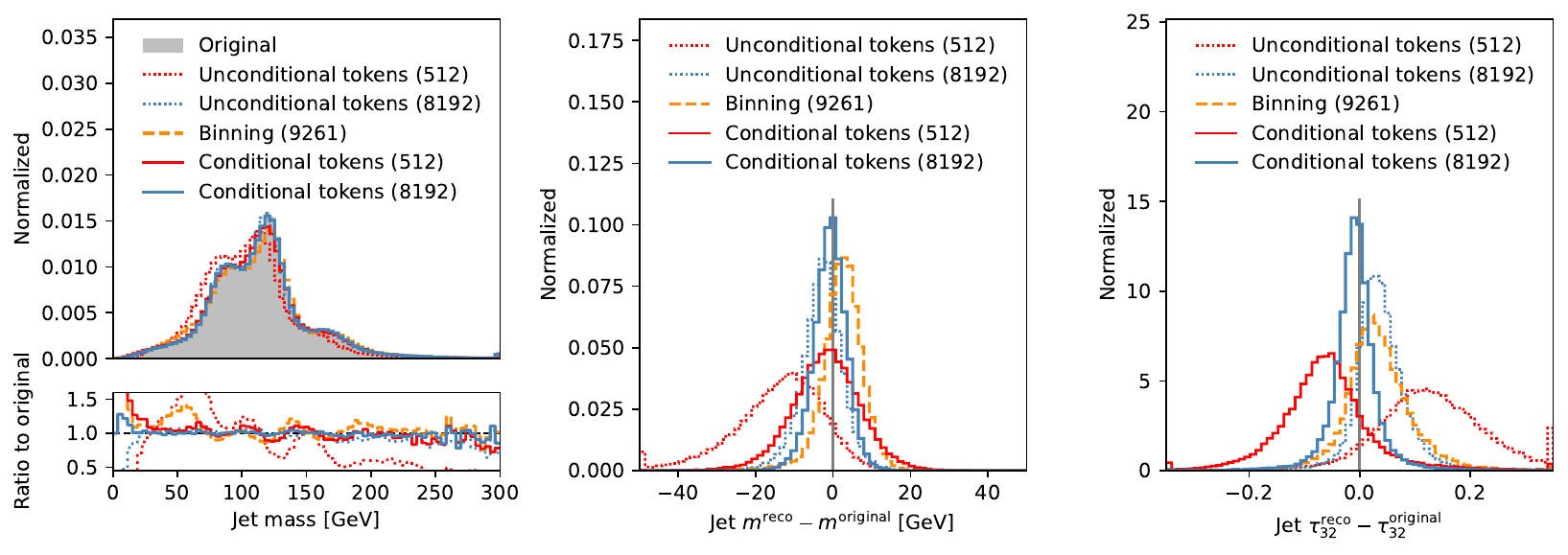}
	\caption{
		(left) Jet mass distribution for all ten jet types combined.
		(center) Difference between the mass after tokenization and the initial mass for \thad\ jets.
		(right) Difference between the $\tau_{32}$ of the initial jets
		and the reconstructed jets for \thad\ jets.
	}
	\label{fig:tokenization_approach_comparison}
\end{figure*}

We first inspect how well the \textbf{tokens cover the space}.
An illustration of the conditional and unconditional tokens in physical space
(i.e. their corresponding \etarel, \phirel\ and \pt\ values)
is shown in \autoref{fig:token_visualization} for the different tokenization
approaches and different codebook sizes.
In the unconditional case, as well as in the binning approach, the
reconstruction of a token is always the same, independent of the other tokens in
the jet, leading to discrete  points in physical space.
In the conditional case, however, the reconstruction of a token is by construction
affected by the other tokens inside this jet.
To visualize the spread of each conditional token in physical space, we reconstruct
each token 500 times conditioned on 50 randomly chosen tokens.
Each of those reconstructions is shown in the scatter plots in
\autoref{fig:token_visualization}, where the different reconstructions of the same
token are drawn in the same color.
We notice that the reconstruction of each token only changes slightly when
conditioned on other tokens.
Thus, the 500 different reconstructions of a conditional token show up in
\autoref{fig:token_visualization} as a blob in physical space.
This already shows that the conditional tokenization allows to cover a
significantly larger volume in reconstruction space, while the unconditional
tokens can only be reconstructed to distinct points in reconstruction space.
We note that our approach of reconstructing each token 500 times conditioned
of randomly chosen other tokens not necessarily represents the reconstructed
values of actual jet constituents, as it is possible that those combinations
of tokens would not appear for real jets.
However, this illustrates the overall behavior of how much the reconstruction
of a token can change due to the conditioning on the other tokens.

Next, we consider \textbf{distributions at the jet level}
to judge the quality of the tokenization.
For this and the following studies, jets are mapped into token space, and then
mapped back to physical space to quantify the loss in information.
\autoref{fig:tokenization_approach_comparison}~(left) shows the jet mass
combined for all classes in the dataset, as was done in~\cite{heinrich2024masked}.
We observe a slight slope in the ratio plot for the unconditional tokens with 
codebook size 512, which is due to the underestimation of the masses (see also fig 8). 
The other approaches have a ratio close to 1 across the bulk of the histogram, 
and only show deviations at the tails where the statistics is low.

\begin{figure}
	\centering
	\includegraphics[width=0.9\linewidth]{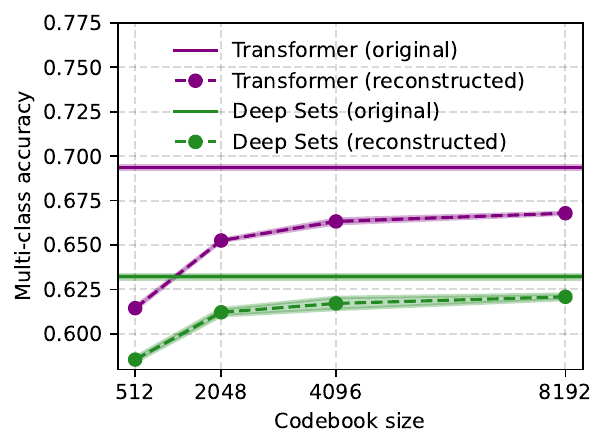}
	\caption{
		Token quality evaluation using a multi-class classifier. The classifier
		accuracy is shown for different codebook sizes and different classifier
		architectures (purple and green).
		The classifiers are also trained on the original constituents, showing
		an upper limit for the achievable accuracy.
		The reconstructed constituents are obtained using the conditional tokenization.
		The reported values and the uncertainty band correspond to the mean and 
		standard deviation over 5 trainings with different random seeds for the 
		randomly initialized weights of the classifier.
	}
	\label{fig:classifier_token_quality_plot}
\end{figure}

However, this inclusive distribution might hide differences at the level of
individual classes and jets. We therefore also consider the difference in mass
for jets before and after tokenization and reconstruction
\autoref{fig:tokenization_approach_comparison}~(center) for \thad\ jets.
The unconditional tokenization leads to a systematic shift of approximately
15~GeV, while the conditional tokenization is well centered at zero.
Increasing the codebook size from 512 to 8192 tokens substantially improves the
resolution.
This behavior is even more pronounced when considering the $N$-subjettiness~\cite{Thaler:2010tr} 
ratio $\tau_{32}$ in \autoref{fig:tokenization_approach_comparison}~(right). 
Both conditional and unconditional tokenization with 512 tokens results in
drastically shifted distributions, while the larger codebook size of 8192
recovers a peak close to zero for the mass difference. Furthermore, using 8192
tokens moves the peak of the $\tau_{32}$ difference closer to zero, while a small bias
towards smaller $\tau_{32}$ in reconstructed jets is still present.
In addition to that, while the mass resolution of the conditional tokens is already
centered close to zero when using a codebook size of 512, the width of the distribution
improves drastically from $\sigma_{512}^\mathrm{cond}=8.3$~GeV to $\sigma_{8192}^\mathrm{cond}=4.0$~GeV 
when moving from a codebook size of 512 to a codebook size of 8192, where $\sigma$ corresponds
to the standard deviation obtained from fitting a normal distribution to the mass
resolution histograms.
A similar behavior can be observed for other classes, where in
some cases, depending on the jet observable and the jet type, the effect is even more extreme. 
Finally, the binning approach with a 21x21x21 linear bins\footnote{
	The binning approach with 9261 tokens in a 21x21x21 grid is used as a reference, 
	as it is the closest (but larger) number of bins (obtained with a NxNxN grid) 
	compared to the codebook size of 8192 tokens.
}
in the input features of our
VQ-VAE comes close to the mass resolution of the conditional tokens with a codebook size
of 8192, while the resolution of the subjettiness ratio is notably worse.
Moreover, while this binning approach with 9162 tokens leads to reasonable resolution
of the \thad\ jets shown in \autoref{fig:tokenization_approach_comparison}, we found
quite drastic mismodeling for \tlep\ and \hlnuqq\ jets with such small codebook
sizes\footnote{
    As expected, the resolution of the binning approach automatically leads to good
    resolution when the codebook size (i.e. the number of bins) is increased 
    to a sufficiently large number. We found that around \num{64000} tokens
    (corresponding to a 40x40x40 grid) offer similar resolution as
    conditional tokenization with a codebook size of 8192.
}. 
The distributions and the corresponding resolutions of the jet mass, jet \pt, as well
as the subjettiness ratios $\tau_{32}$ and $\tau_{21}$ are shown for all ten jet
types individually in \autoref{sec:token_quality_appendix}.
Overall, the highest fidelity is achieved by conditional tokenization with a marked
improvement from increasing the codebook size from 512 to 8192.

Finally, we \textbf{quantify the information loss that comes with the tokenization by training multi-class classifiers} to distinguish between
the ten jet types present in the dataset.
The classifiers are trained once with the original inputs, and once with the inputs
after undergoing tokenization and subsequent reconstruction back into physical space.
This procedure allows a direct comparison of how the loss in resolution affects
reconstruction performance.
We utilize two standard classifier
architectures: Deep~Sets~\cite{zaheer2018deep,Komiske_2019}
(i.e. without particle interactions) and Transformer~\cite{Vaswani:2017lxt,shleifer2021normformer}
(i.e. with particle interactions) and perform this study for four different
codebook sizes from 512 to 8192 tokens for the conditional tokenization
approach.
Details about the classifier trainings can be found in
\autoref{subsec:token_quality_classifier_details_appendix}.
Note that this approach is similar in spirit to the classifier metric proposed
in~\cite{Krause:2021ilc,understandLimitationsOfGen}
but tests a multi-class
classifier trained on these samples individually, as opposed to judging how well
a classifier might distinguish original and reconstructed samples. This is
necessary as e.g. points at fixed positions would be distinguishable from the
original with close to perfect accuracy, rendering the test less useful.

The resulting classifier accuracy for the two different architectures is shown
in \autoref{fig:classifier_token_quality_plot}.
As seen in previous studies of resolution, we observe
an increase of token quality as we increase the size of the VQ-VAE codebook.
Furthermore, we see that the classifier performance starts to plateau with codebook sizes
larger than 4096. However, even at the largest codebook size, a gap to the
performance on original particles remains, motivating future work into building
more accurate tokenization schemes.

For the remaining studies we will utilize a codebook size of 8192 with
conditional tokens as this leads to the overall best performance and fidelity.

\begin{figure*}
	\centering
	\includegraphics[scale=0.32]{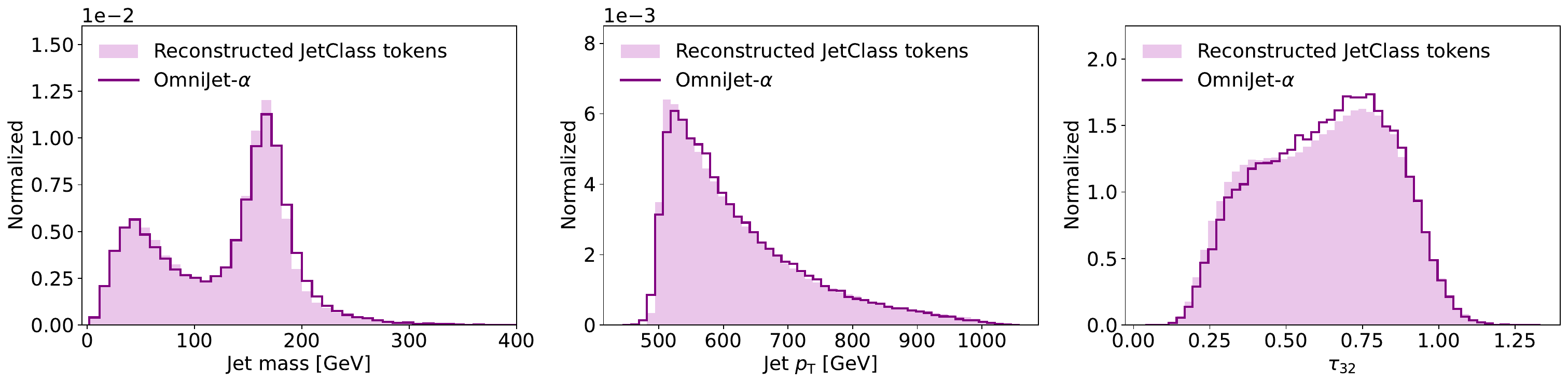}
	\includegraphics[scale=0.32]{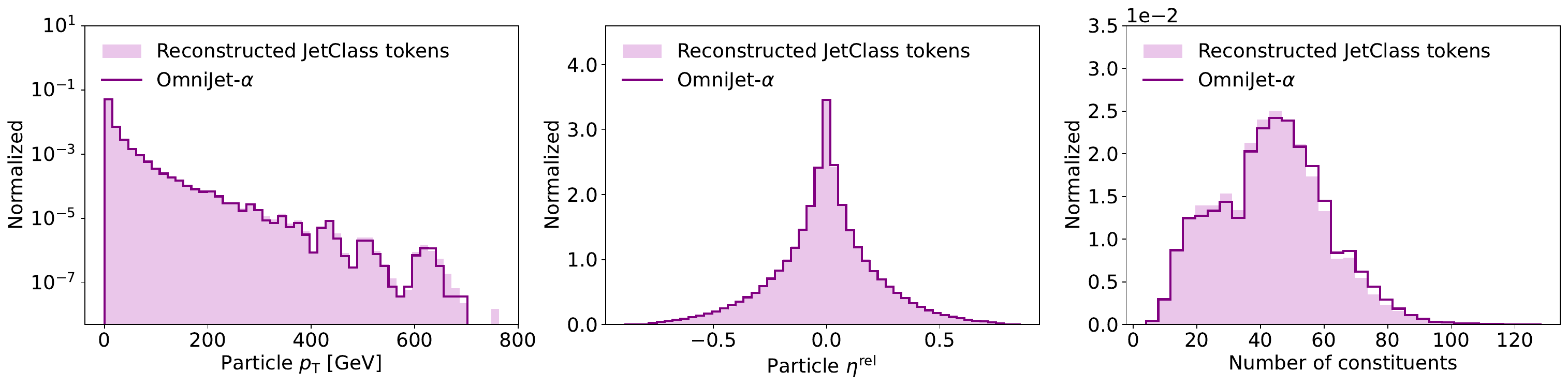}
	\caption{
            Comparison of generated jets from the model trained on both \qcd{} and
            \thad{} jets, to reconstructed \jetclass{} tokens. The top row shows jet
            level distributions, while the bottom row shows distributions on the
            constituent level.
 }
	\label{fig:gen_vs_truth}
\end{figure*}

\subsection{Jet generation}
After training the transformer backbone with the generative head, it can be used for autoregressive generation as described in \autoref{sec:model}. The model was trained on three separate datasets: \thad{} only, \qcd{} only, and \qcd{} and \thad{} combined. This section will describe the combined version, since this is the
model that is used for transfer learning. For a discussion of single-jet type generative results, including a comparison to the EPiC-FM model of~\cite{Birk:2023efj}, see appendix \autoref{sec:more_plots}. 48\,000 events were generated
from the combined model. These events contain tokens, which are then decoded back to physical space using the VQ-VAE decoder.

A comparison to reconstructed \jetclass\ tokens can be seen in  \autoref{fig:gen_vs_truth}. We observe that in general the
model is able to match the truth level tokens well. 
We note that the tail of the $p_\mathrm{T}$ spectrum of both the generated constituents and the reconstructed \jetclass\ tokens contains bumps distributed around
discrete values, which is consistent with our inspection of the reconstruction space shown in \autoref{fig:token_visualization}.

In order to quantify the performance, a classifier test (see \autoref{subsec:generation_classifier_tests} for details) is performed to distinguish generated events from reconstructed \jetclass{} tokens. The test results in an AUC score of 0.54.

\subsection{Transfer learning from generation to classification}
\label{subsec:transfer_learning_results}

\begin{figure*}
	\centering
	\includegraphics[width=0.97\linewidth]{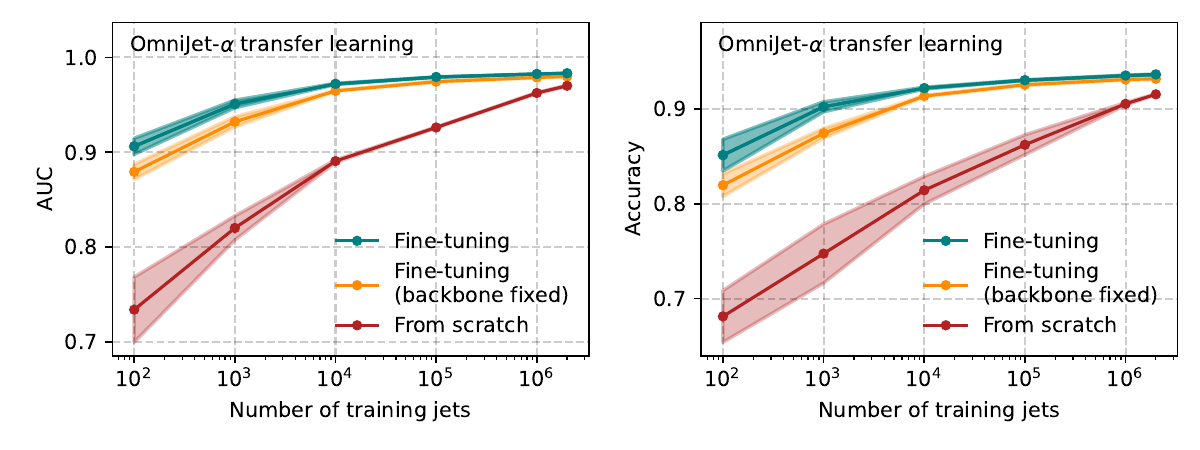}
	\caption{
		Performance of pre-trained and non-pre-trained models for the task of \thad\
		vs \qcd\ jet classification. The area under the ROC-curve (AUC) metric is
		shown on the left, the classification accuracy on the right.
		For each training strategy and dataset size, the mean and standard deviation
		over 5 trainings are shown, where the 5 trainings are performed with different
		random seeds for the randomly initialized weights of the classifier.
		For the pre-trained models, the same backbone weights are used for all 5 trainings.
	}
	\label{fig:transfer_learning_acc_and_auc_vs_dataset_size}
\end{figure*}

To evaluate the ability of the model to generalize from generating jets to
classifying them, we focus on the task of hadronic top quark
tagging~\cite{Kasieczka:2017nvn,Kasieczka_2019}, i.e. distinguishing \thad\ and \qcd\ jets on the
\jetclass~\cite{JetClass} dataset.
For this test, the \textit{Next-token prediction head} is replaced by a
\textit{Classification head} while the transformer backbone is
retained.
We compare three training strategies:
training the full architecture with randomly initialized weights (termed
\textit{from scratch}) which does not use transfer-learning and corresponds to
the baseline, and two versions of fine-tuning the model obtained in the
generative training step.
In the \textit{Fine-tuning} run, both the pre-trained backbone weights and the
randomly initialized classification task head weights are allowed to float in
the training, while in \textit{Fine-tuning (backbone fixed)} only the
classification task head is allowed to change.

The results of these training runs are presented in
\autoref{fig:transfer_learning_acc_and_auc_vs_dataset_size} as a function of
the number of training examples provided to the model.
We observe a significant gain in classification accuracy of both fine-tuning
approaches compared to the baseline, leading to up to 17 percentage-points
higher accuracy for small number of training jets, and outperforming by a few
percentage-points at the highest training sample size. The difference between
the two fine-tuning strategies is relatively small, with the more open training
performing slightly better.
Put differently, the generative pre-trained model achieves an accuracy of
around 85\% 
with 100 training examples for which the model that is trained from scratch requires more than 10\,000 examples.

\section{Conclusion}
\label{sec:conclusion}

Foundation models for physics data are an enticing promise: Trained on large
amounts of data and tasks, they are expected to easily generalize to any
down-stream problem, saving countless hours of human and compute time.
In this paper we have taken crucial steps towards the creation of such models.

First, we expect learned representations of data to play a key role as inputs to
foundation models. Representations might be continuous and rely on
symmetries
or learn a mapping to a
discrete space as done here with tokenization. Note that while using data
\textit{raw} --- i.e. without prior mapping into a representation space ---
might be possible when only considering a narrow range of similar datasets, it
is inherently limiting when data from different sources or with different
initial dimensionalities are to be considered.

Whatever representation is used, it will be important to understand and minimize
the loss of information inherent in this transformation.
This problem is especially important for downstream uses such as classification
and regression tasks, as the loss of information can directly limit the
achievable accuracy or resolution.
This work introduced a set of criteria --- both distribution and classifier
based --- that can be used to assess the quality of any representation.

Using these metrics, we found a marked increase in the resolution of relevant
observables like mass and jet substructure by using a codebook size of 8192 with
conditional tokenization over binning-based approaches, unconditional
tokenization, and conditional tokenization with smaller codebooks. An additional
classifier test further confirmed this observation.

Next, we demonstrated the training of an autoregressive generative model for jet
constituents, specifically for \qcd\ and \thad\ jets from the
\jetclass\  dataset~\cite{JetClass}.
The generated distributions agree well with the ground truth, both for global
jet kinematics, jet substructure, and individual constituent features.
We note that while our model is the first token-based generator of \jetclass -like
examples, more extensive studies of its generative fidelity when increasing the
feature-space and detailed comparison to prior non-token-based results on this
dataset~\cite{Birk:2023efj} are left for future work.

Most importantly, we report the generalization capability of \omnijetalpha{} from
learning to generate jets in an unsupervised way, to the supervised
classification between \thad\ and \qcd\ jets.
Overall, the fine-tuned model outperformed training from scratch for all values
of training examples, often by a significant margin. For example, for 1000
training jets, the fine-tuned model achieves an accuracy of approximately 90\%,
compared to around 75\% for the freshly initialized model. While the two approaches
seem to converge, even at the highest training size of 2 million jets the
fine-tuned approach maintains a lead of a few percentage points. Finally, it
even provides a non-trivial classification accuracy of 85\%, even
when trained only on 100 jets, emphasizing the value of foundation models for
problems with few available labeled training examples. 
Since the backbone trains completely without labels, it allows us to use a large
unlabeled dataset to boost the performance of a smaller labeled one.
While other types of transfer have been demonstrated previously, this is the first time
that the unification across the two \textit{big} classes of tasks --- classification
and generation --- has been achieved.

Of course, this work is only one step in building overarching foundation models.
While it is the first model that achieves both classification and generation, it
is not the most performant for either of these tasks.
However, strategies to increase the performance exist and will be integrated.
For example, the representation quality needs to be improved, possible gains
from masked pre-trained have to be evaluated, architecture and
training data need to be scaled up, and more extensive studies of the
generalization capabilities, including training and testing on additional tasks,
performed. In the medium term, strategies need to be
found to align diverse datasets as well as to integrate pre-trained foundation
models in community workflows.
Nevertheless, the potential benefits in physics performance and compute efficiency
glimpsed at in this and other works makes this a worthy endeavor.

\section*{Code availability}
The code for \omnijetalpha{} is publicly available at \url{https://github.com/uhh-pd-ml/omnijet_alpha}.

\section*{Acknowledgements}

We thank David Shih, Michael Kr\"amer, Michael Kagan,
Frank Gaede, Sarah Heim, and Judith Katzy for stimulating discussions
of foundation models for physics data and Erik Buhmann for valuable comments on the manuscript.

The authors acknowledge support by the Deutsche Forschungsgemeinschaft under
Germany’s Excellence Strategy – EXC 2121  Quantum Universe – 390833306, and
under PUNCH4NFDI – project number 460248186.
This research was supported in part through the Maxwell computational resources
operated at Deutsches Elektronen-Synchrotron DESY, Hamburg, Germany.

\bibliography{refs}

\appendix

\section{Model details and hyperparameters}

Different hyperparameter configurations were tested for the individual model components
of \omnijetalpha{}. The configurations presented in the following were found to lead to 
stable results. 
However, no extensive hyperparameter optimization was performed.

\subsection{VQ-VAE for token creation}
\label{subsec:vqvae_details_appendix}

Both the \etarel\ and the \phirel\ values are scaled down by a factor of~3.
The transverse momentum of the jet constituents is first transformed using the
natural logarithm and subsequently shifted by -1.8.
The tokenization was also done without the log transform of the \pt, and was found to
perform similarly.
However, the logarithm transformation has the advantage that it automatically avoids
negative \pt\ values, which is why we choose to use the log-transformed \pt.
The conditional and unconditional tokenization only differ in the architecture
of the encoder and decoder of the VQ-VAE.

Training for the VQ-VAE is implemented in \texttt{pytorch}~\cite{pytorch_NEURIPS2019_9015}
and \texttt{pytorch-lightning}~\cite{falcon_2024_10779019}.

The model architecture of the VQ-VAE encoder and decoder in the conditional tokenization approach is similar 
to~\cite{heinrich2024masked} with a different set of hyperparameters.
We use 4 NormFormer~\cite{shleifer2021normformer} blocks with an embedding
dimension of 128 and 8 heads in the MHA for both the encoder  and the decoder.
We use the \texttt{vqtorch} library~\cite{huh2023vqtorch,
huh2023straightening} to implement the vector quantization layer with the dimension of the
codebook vectors set to 4.

The mean squared error (MSE) between the tensor of the initial particle features
and the reconstructed features is used as the task loss
$\mathcal{L}_\mathrm{task}$.
The total loss is then set to 
\begin{align}
    \mathcal{L} = \mathcal{L}_\mathrm{task} + \alpha \cdot \mathcal{L}_\mathrm{commit} 
\end{align}
with  $\alpha=10$.
An affine transformation is used for a joint transformation of all codes 
and dead codes are replaced with a frequency of 10 iterations.
The parameter $\beta$ which trades off the importance of updating the
embeddings from the encoder $z_e$ and the code vectors $z_q$
is set to $\beta=0.9$. Lastly, we use a synchronized update 
rule~\cite{huh2023straightening,huh2023vqtorch} with $\nu=1$.

In the unconditional approach, we use the same hyperparameters as outlined above, with
the only difference that the architecture of the encoder and decoder is a simple MLP with 
3 hidden layers of dimension 128 and ReLU activation function.

All VQ-VAE models are trained on all 10 classes of the \jetclass{} dataset~\cite{JetClass}.

\subsection{Classifiers for token quality evaluation}
\label{subsec:token_quality_classifier_details_appendix}

The DeepSets~\cite{zaheer2018deep,Komiske_2019} classifier consists of a
per-particle MLP $\Phi$ with shared weights across all particles inside
the jet with 3 hidden layers of dimension 100, 100 and 256. 
The output of the network $\Phi$ is then aggregated with a sum and
fed into another MLP with 3 hidden layers of dimension 100 followed
by a 10-dimension output layer with softmax activation function.

The Transformer classifier consists of 5 NormFormer~\cite{shleifer2021normformer}
blocks, followed by two class-attention blocks with a class token as query,
inspired by the ParT~\cite{qu2024particle} architecture. 
The output of the last class-attention block is fed into a MLP with two
hidden layers of dimension 128, followed by a softmaxed 10-dimensional
output layer.

The classifiers are trained with the AdamW~\cite{AdamW} optimizer with a maximum
learning rate of 0.005 (0.001) for the DeepSets (Transformer) classifier
and weight decay 0.01.
The learning rate first linearly increased from 0.002 (0.0005) during the first
4 training epochs, after which it is linearly decreased to the initial learning rate
over 20 epochs and then linearly decreased to a final learning rate of 0.001 (0.0003),
following the one-cycle learning rate schedule \cite{smith2018disciplined}.

The classifiers for those token quality evaluations are trained on 10M jets
from the \jetclass\ dataset~\cite{JetClass}.

\subsection{Transformer backbone}

When training the transformer backbone, cross entropy is used as a loss function and Adam~\cite{kingma2017adam} with a learning
rate of 0.001 as optimizer. The model had access to 10M \thad{} jet events and 10M \qcd{} jet events. Note that this means that the model trained on these two jet types combined had access to twice as much data. All versions were trained for 30 epochs, and the model state with the lowest validation loss was chosen for the further analysis.

\subsection{Transfer learning}
\label{subsec:transfer_learning_details_appendix}

To perform the transfer learning from the generative task to the
classification task, we change the head of the \omnijetalpha\ model
to the classification head and load the weights of the backbone trained
for the generative task.
We explore two variations of fine-tuning the pre-trained backbone to the classification
task: training all weights of the model with the same learning rate (referred to
as \textit{Fine-tuning} in \autoref{subsec:transfer_learning_results}) and keeping
the weights of the backbone fixed at the state obtained from the generative model
(referred to as \textit{Fine-tuning (backbone fixed)} in
\autoref{subsec:transfer_learning_results}). For the \textit{From scratch} trainings
we start the training with randomly initialized weights of the whole model.
The training is performed with the AdamW~\cite{AdamW} optimizer with a constant
learning rate of 0.0001 and weight decay 0.01.
The transfer learning studies are performed on the test dataset from the default
train-val-test split of the JetClass dataset, which includes 2M jets per jet-type,
leading to 4M jets in total for our studies with \qcd{} and \thad{} jets.
We split those 4M jets into 2M jets for training, and 1M jets for validation and
testing, respectively.
Since those trainings, depending on the size of the training dataset, tend to show
overfitting quite quickly, we stop those trainings when the validation loss does not
improve for multiple epochs. The threshold of this early stopping is adjusted to
the training dataset size, with a patience of 20 epochs for a training dataset of
\num{100}, \num{1000} and \num{10000} jets, a patience of 10 epochs for a training 
dataset size of \num{100000} and \num{1000000}, and a patience of 5 for trainings with \num{2000000} training jets.
For each training dataset size we run 5 trainings with different random seeds
and the epoch with the smallest validation loss is chosen for evaluation.

\subsection{Classifier tests}
\label{subsec:generation_classifier_tests}
In order to quantify the performance of the generative model, a classifier test using the structure of the DeepSets classifier from \autoref{subsec:token_quality_classifier_details_appendix} is performed. In this case however, the 3 hidden layers of the particle MLP $\Phi$ all have dimension 10.

A number of 48\,000 generated events are combined with equally many reconstructed tokens from the test set of \jetclass{}. The two datasets are combined and shuffled, and a train/val/test split of 0.6/0.2/0.2 is used. The model is trained for 100 epochs with binary cross entropy loss and Adam~\cite{kingma2017adam} with learning rate 0.001 as optimizer. The model state with the lowest validation loss is chosen for evaluation. The resulting AUC scores are 0.54 for the model trained on \qcd{} and \thad{} jets combined and 0.57 for the ones trained on single-type jets. 

\section{Token quality}
\label{sec:token_quality_appendix}
\begin{figure*}
	\centering
	\includegraphics[width=0.73\linewidth]{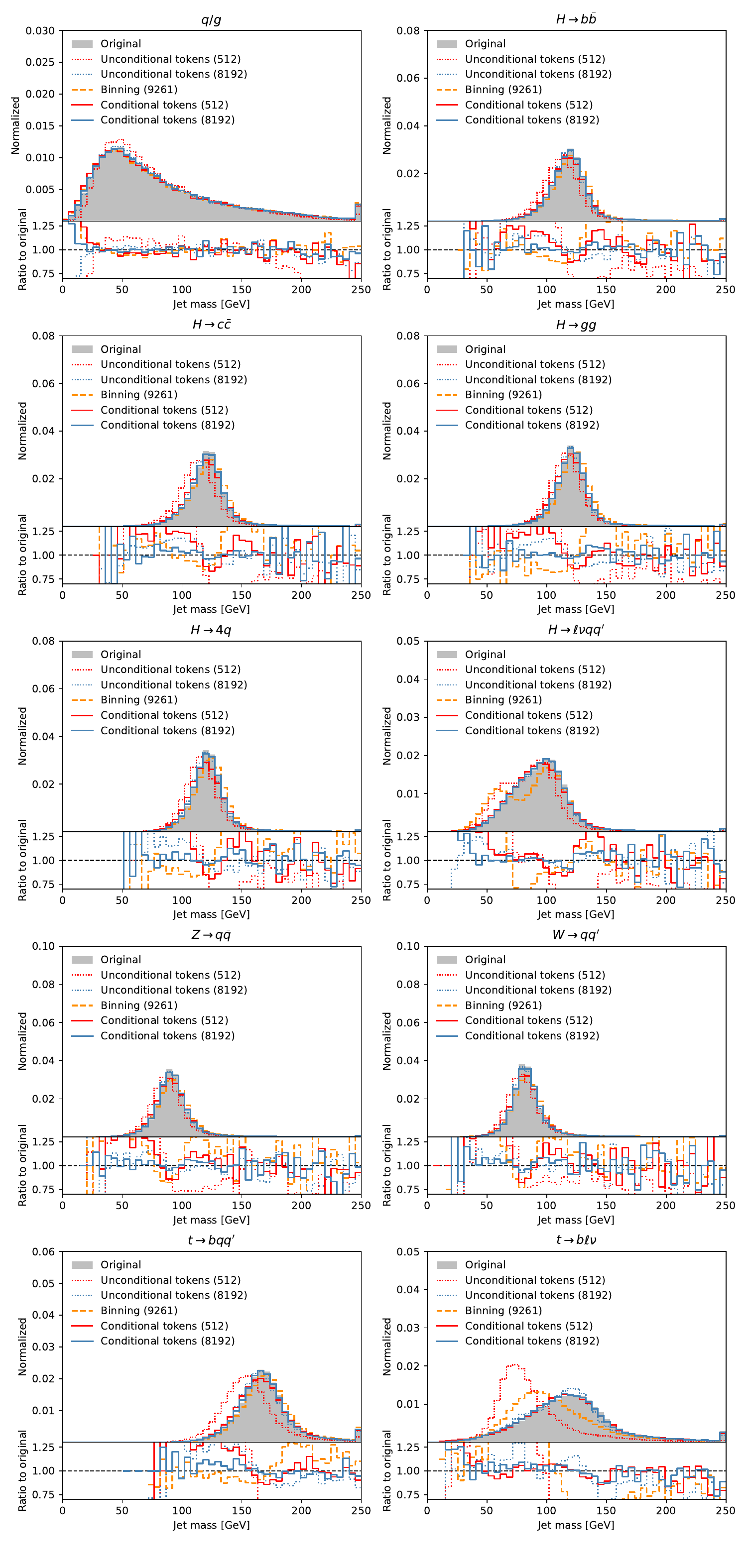}
	\caption{
		Jet mass distribution for different tokenization approaches and codebook sizes.
	}
	\label{fig:jet_mass_distribution}
\end{figure*}

\begin{figure*}
	\centering
	\includegraphics[width=0.85\linewidth]{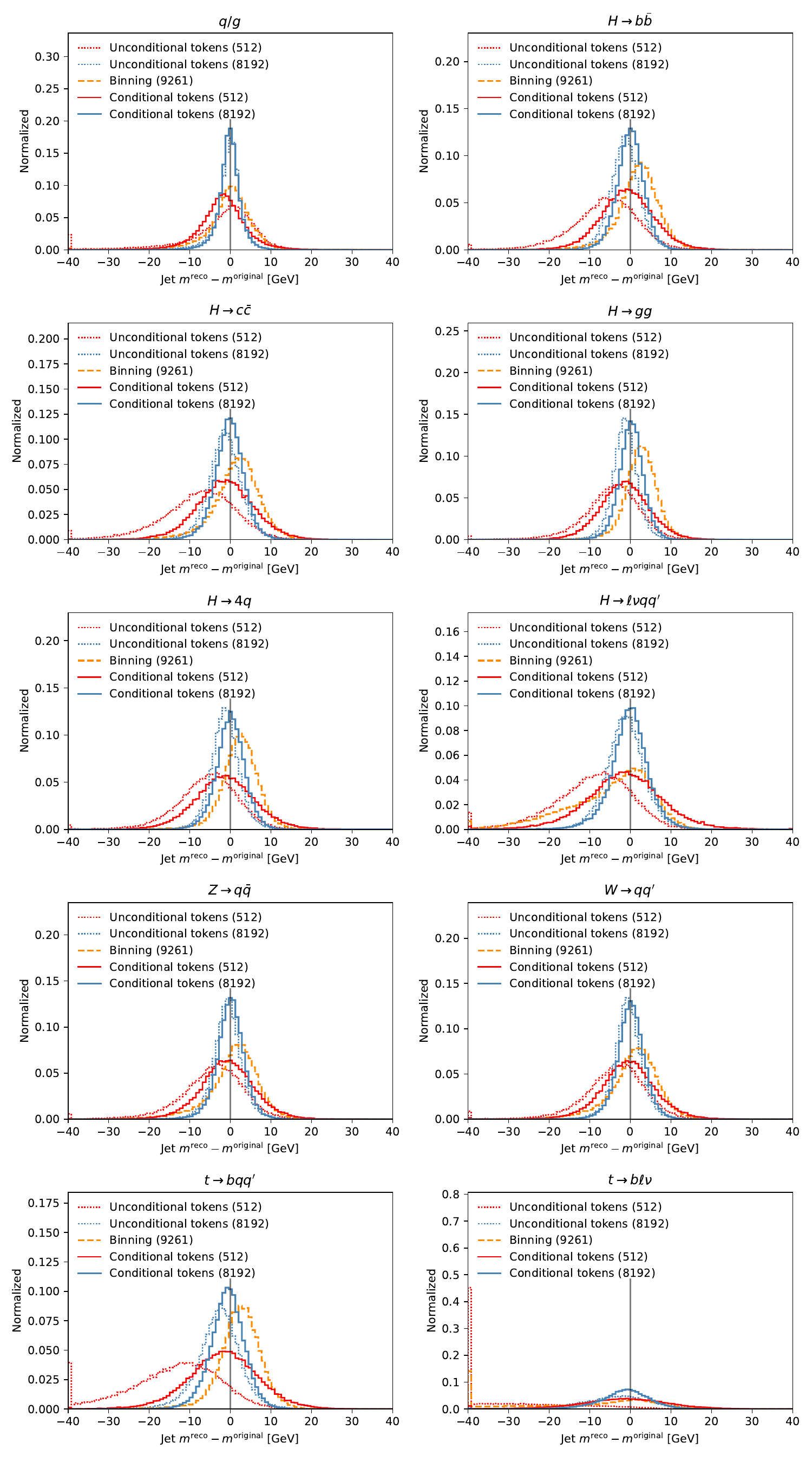}
	\caption{
		Jet mass resolution for different tokenization approaches and codebook sizes.
	}
	\label{fig:jet_mass_resolution}
\end{figure*}

\begin{figure*}
	\centering
	\includegraphics[width=0.73\linewidth]{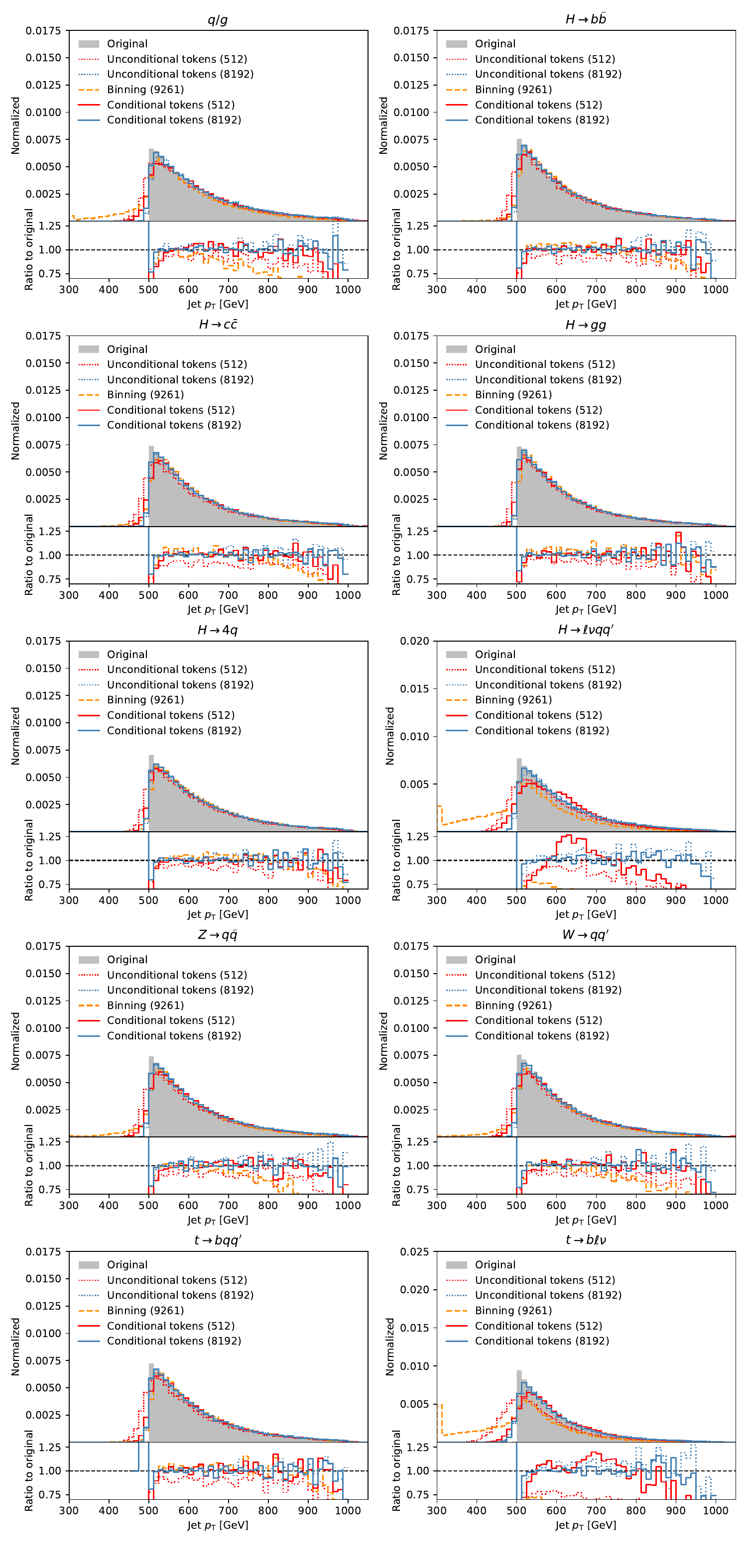}
	\caption{
		Jet \pt\ distribution for different tokenization approaches and codebook sizes.
	}
	\label{fig:jet_pt_distribution}
\end{figure*}

\begin{figure*}
	\centering
	\includegraphics[width=0.85\linewidth]{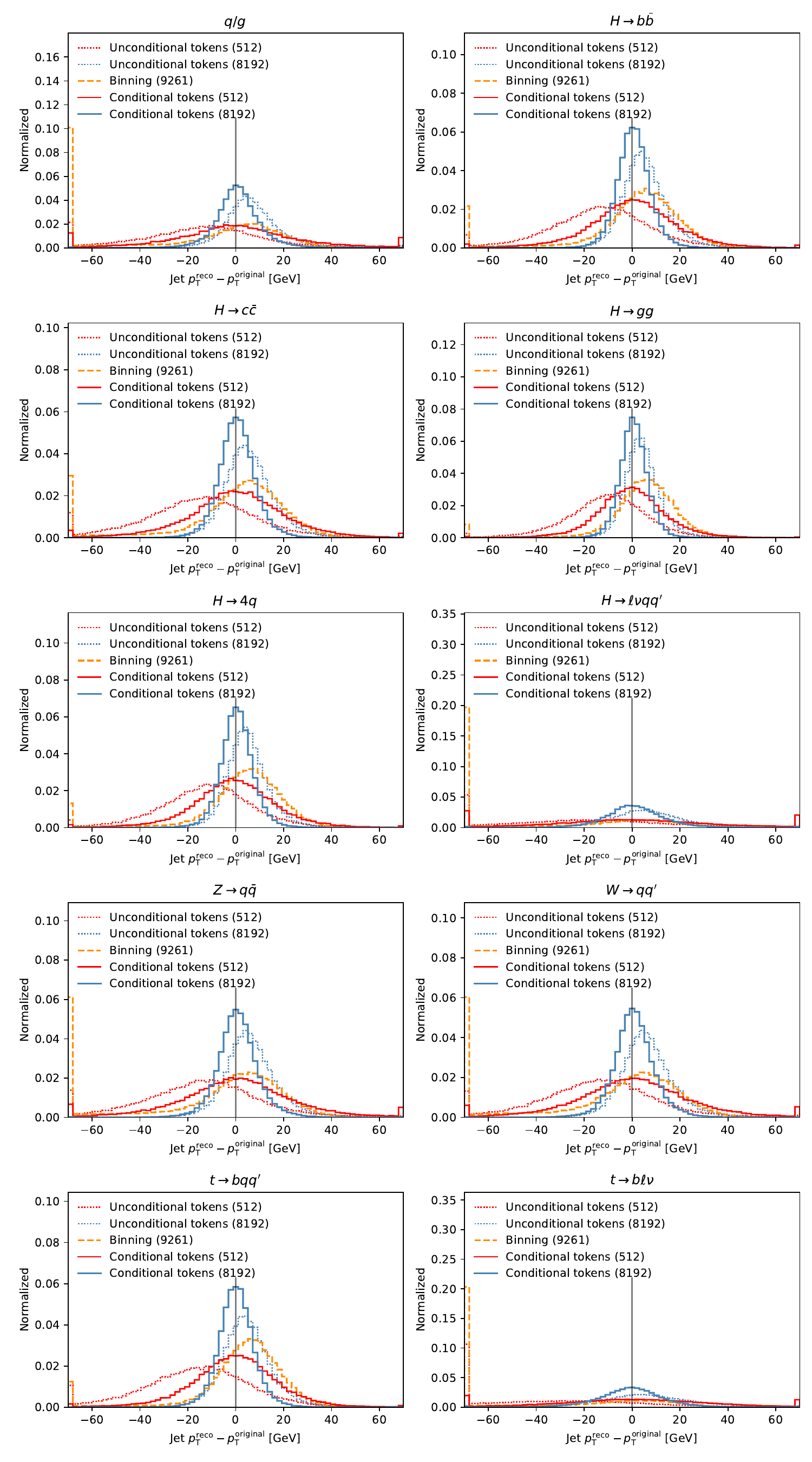}
	\caption{
		Jet \pt\ resolution for different tokenization approaches and codebook sizes.
	}
	\label{fig:jet_pt_resolution}
\end{figure*}

\begin{figure*}
	\centering
	\includegraphics[width=0.73\linewidth]{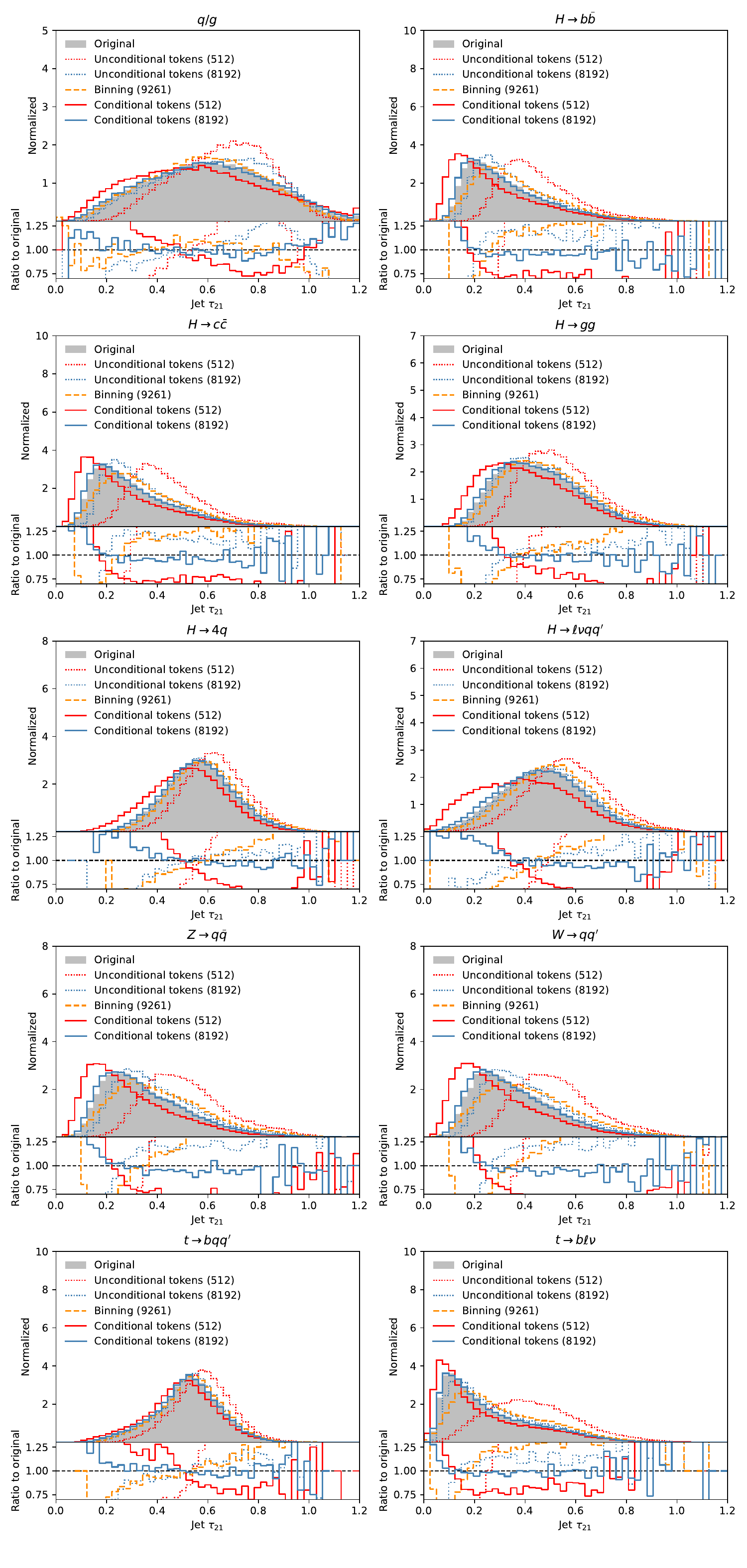}
	\caption{
		Jet $\tau_{21}$ distribution for different tokenization approaches and codebook sizes.
	}
	\label{fig:jet_tau21_distribution}
\end{figure*}

\begin{figure*}
	\centering
	\includegraphics[width=0.85\linewidth]{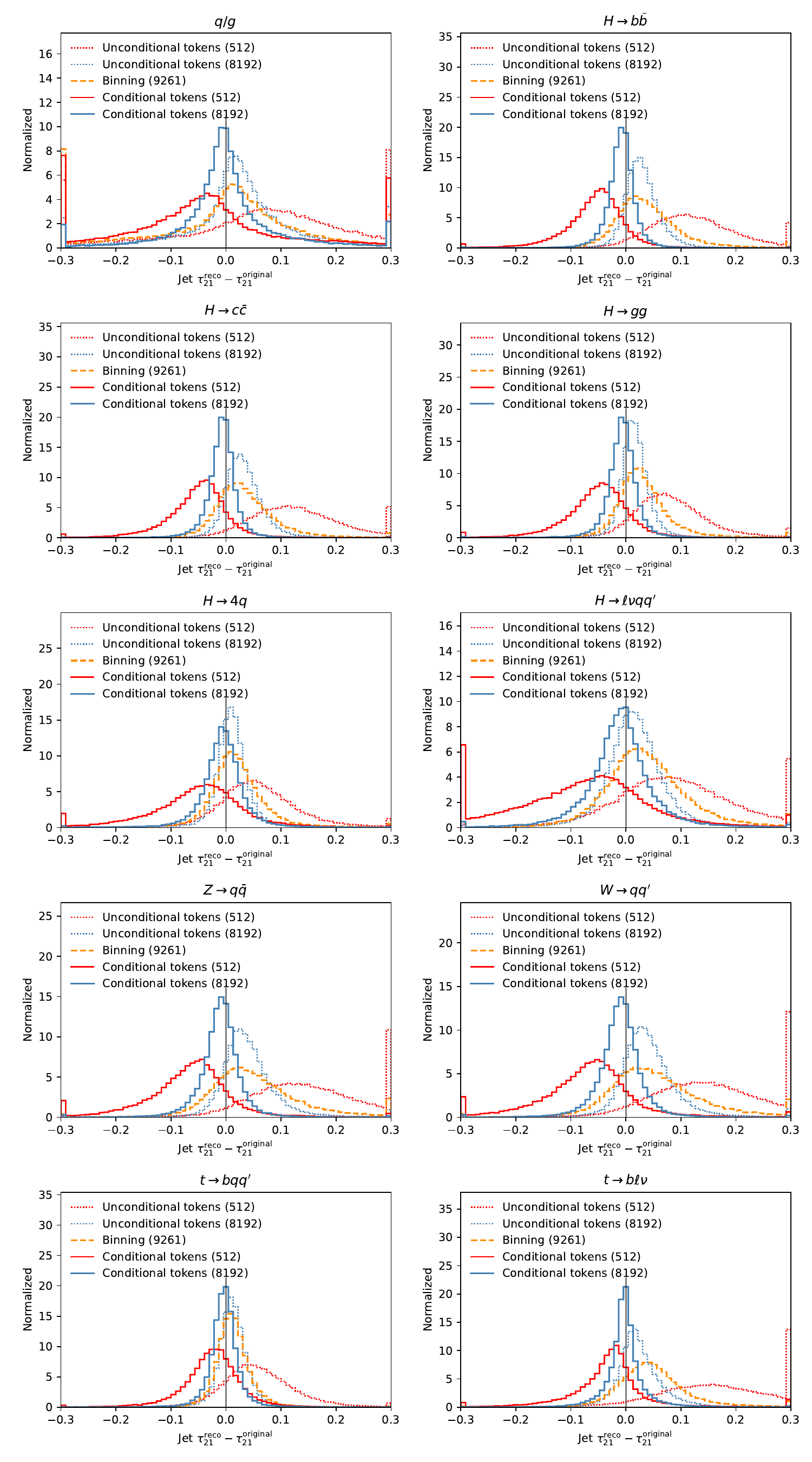}
	\caption{
		Jet $\tau_{21}$ resolution for different tokenization approaches and codebook sizes.
	}
	\label{fig:jet_tau21_resolution}
\end{figure*}

\begin{figure*}
	\centering
	\includegraphics[width=0.73\linewidth]{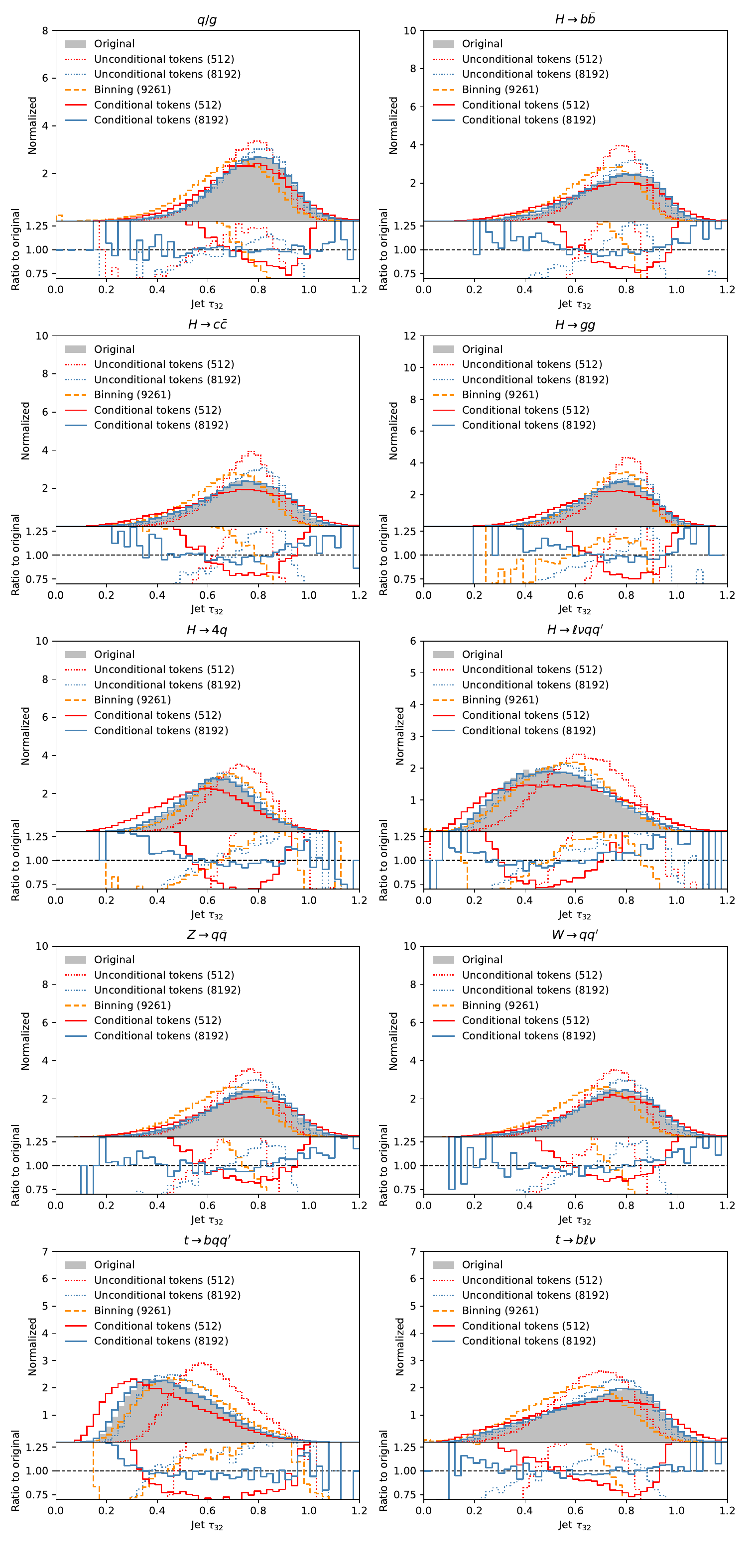}
	\caption{
		Jet $\tau_{32}$ distribution for different tokenization approaches and codebook sizes.
	}
	\label{fig:jet_tau32_distribution}
\end{figure*}

\begin{figure*}
	\centering
	\includegraphics[width=0.85\linewidth]{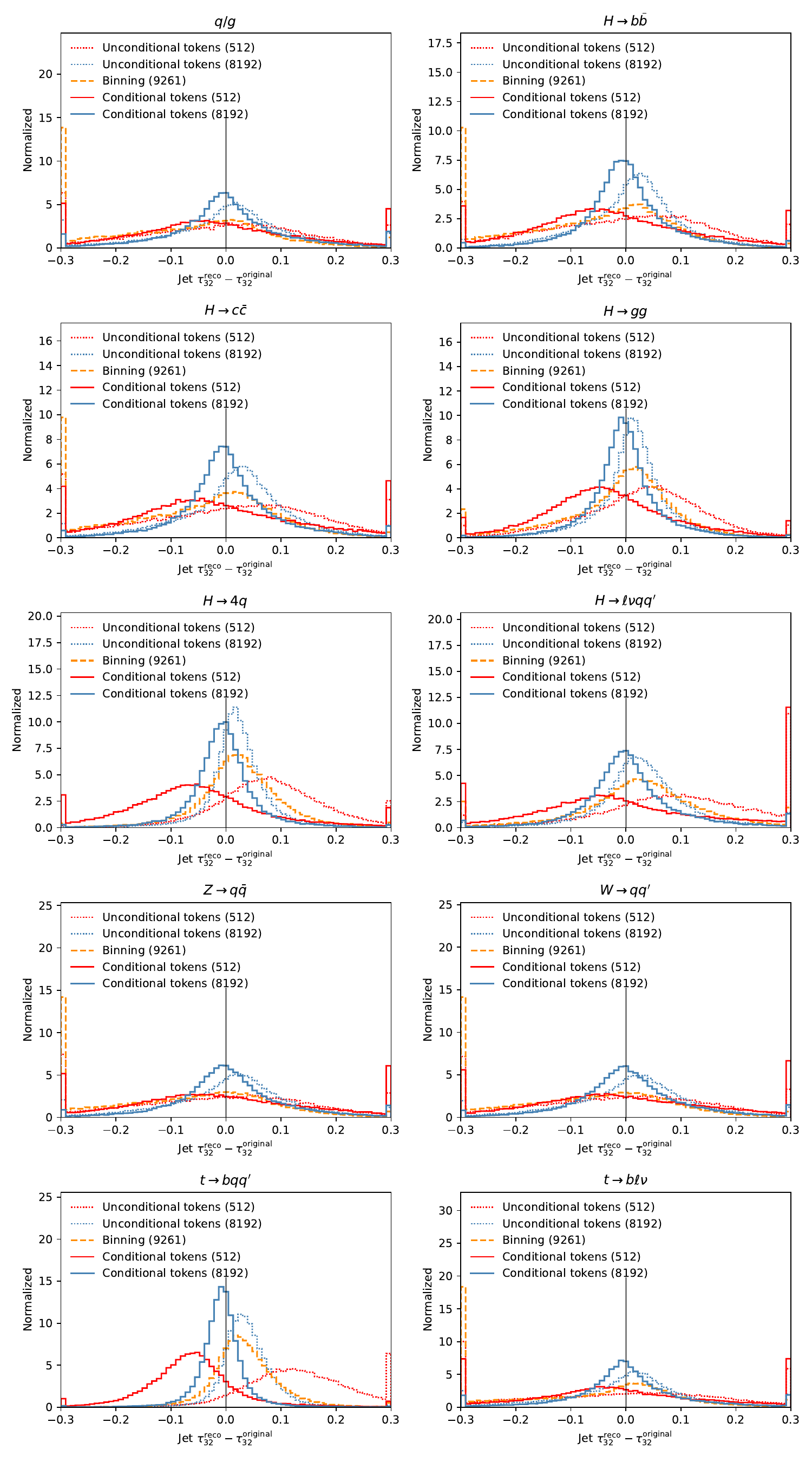}
	\caption{
		Jet $\tau_{32}$ resolution for different tokenization approaches and codebook sizes.
	}
	\label{fig:jet_tau32_resolution}
\end{figure*}

Additional plots of the jet mass, the jet transverse momentum, as well as the subjettiness
ratios are shown in 
\autoref{fig:jet_mass_distribution}, \autoref{fig:jet_mass_resolution}, 
\autoref{fig:jet_pt_distribution}, \autoref{fig:jet_pt_resolution}, 
\autoref{fig:jet_tau21_distribution}, \autoref{fig:jet_tau21_resolution},
\autoref{fig:jet_tau32_distribution} and \autoref{fig:jet_tau32_resolution}.

\section{Generative model trained on single-jet data}

\label{sec:more_plots}

\begin{figure*}
	\centering
	\includegraphics[scale=0.32]{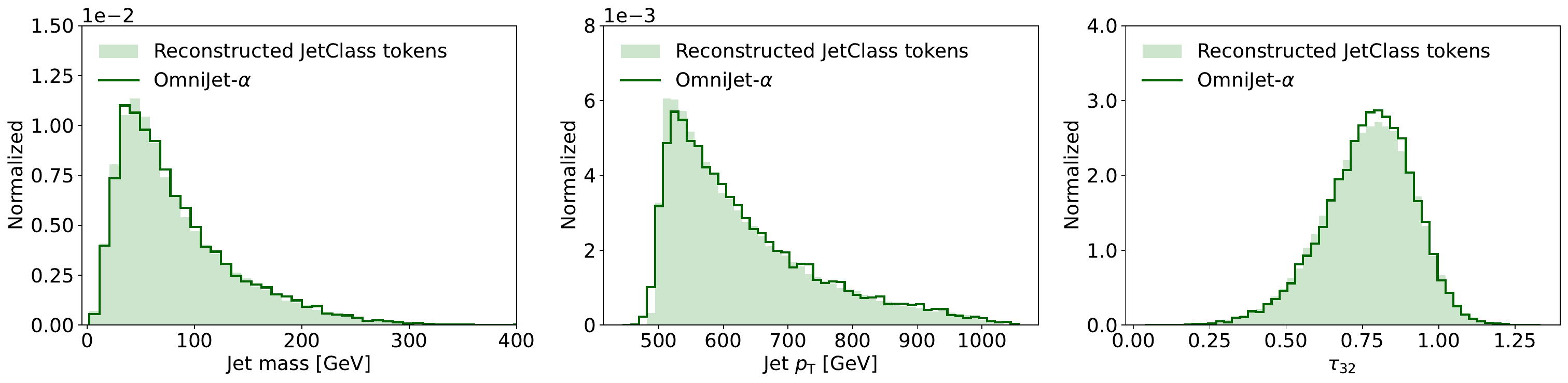}
	\includegraphics[scale=0.32]{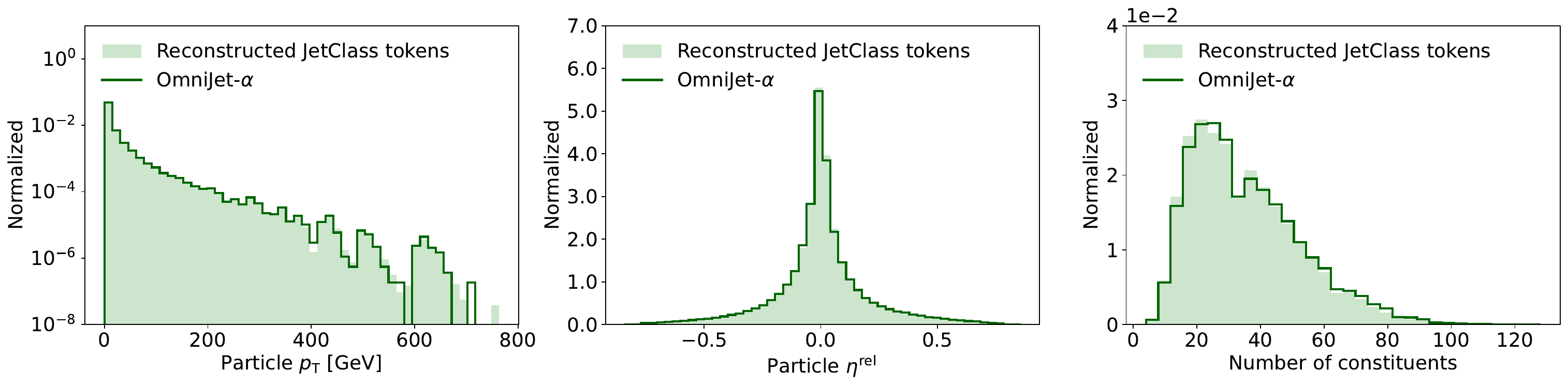}
	\caption{Comparison of generated jets from the model       trained on \qcd{} jets to reconstructed \jetclass{} tokens. The top row shows jet
            level distributions, while the bottom row shows distributions on the
            constituent level.}
	\label{fig:gen_vs_truth_QCD}
\end{figure*}

\begin{figure*}
	\centering
	\includegraphics[scale=0.32]{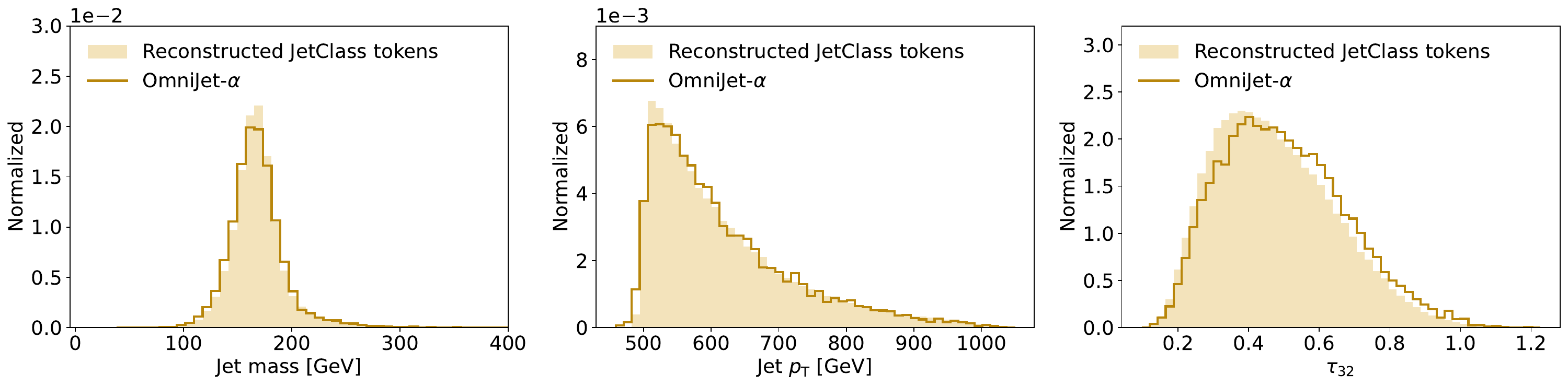}
	\includegraphics[scale=0.32]{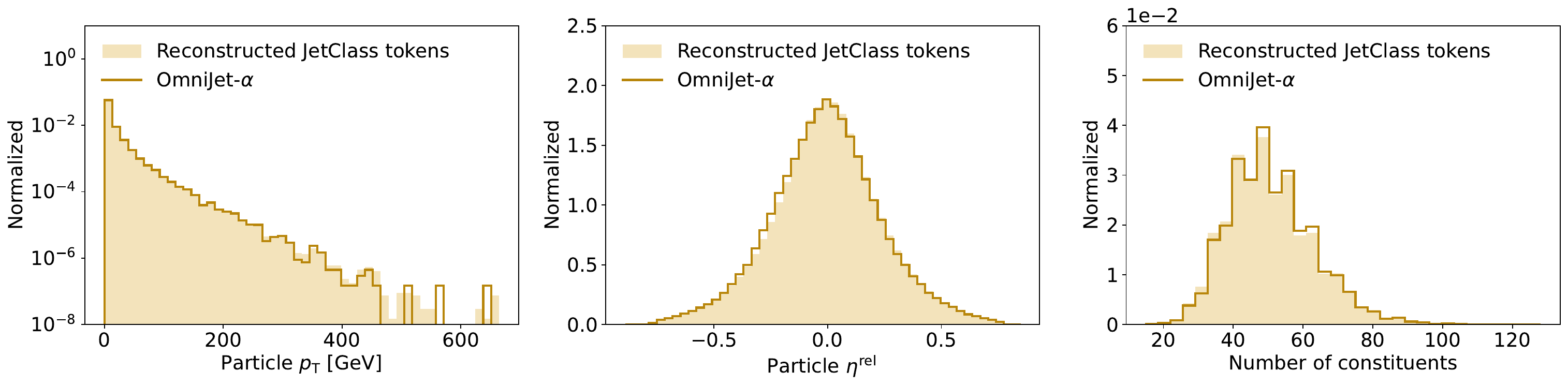}
	\caption{Comparison of generated jets from the model trained on \thad{} jets to reconstructed \jetclass{} tokens. The top row shows jet
            level distributions, while the bottom row shows distributions on the
            constituent level.}
	\label{fig:gen_vs_truth_top}
\end{figure*}

\begin{figure*}
    \centering
    \includegraphics[scale=0.3]{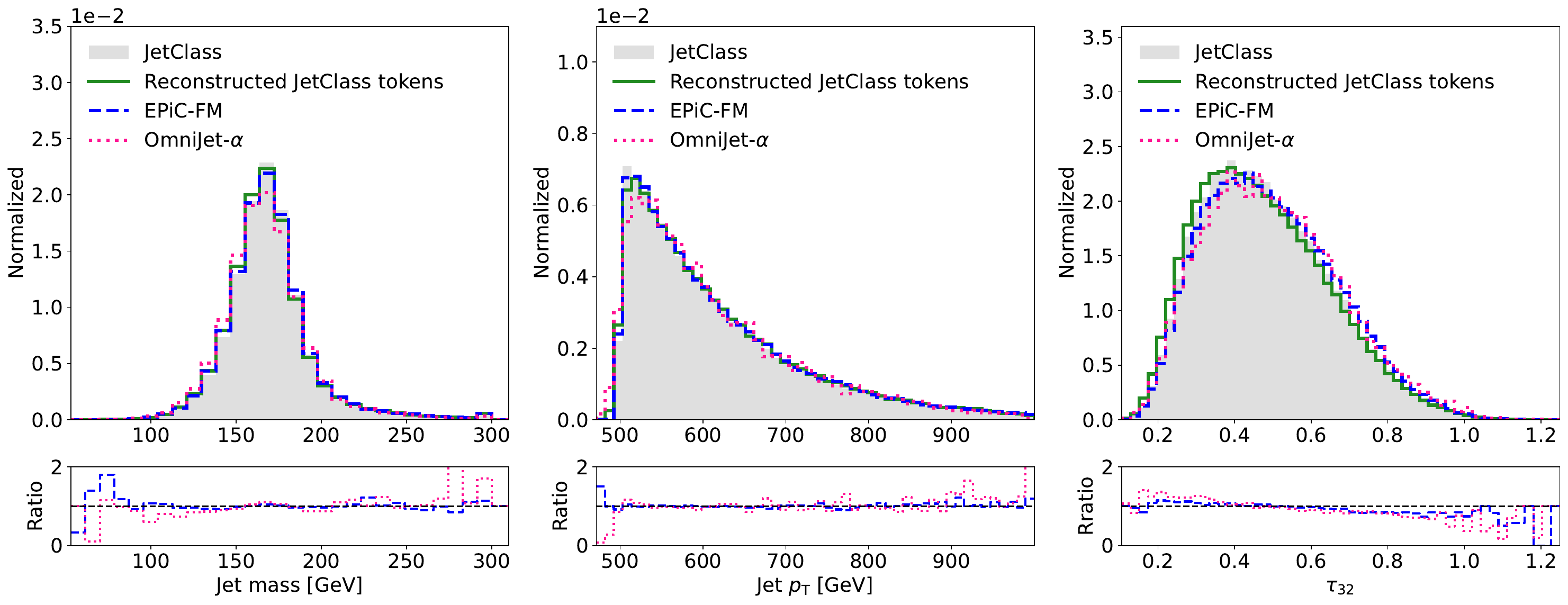}
	\includegraphics[scale=0.3]{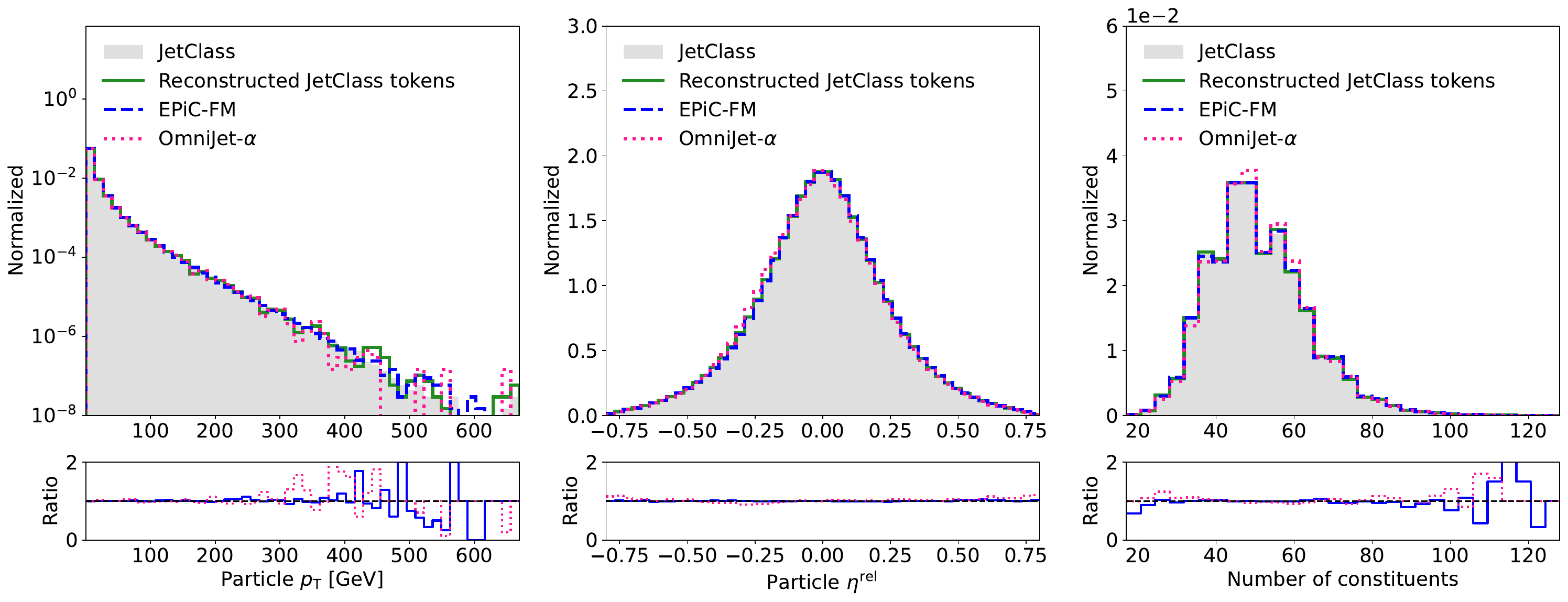}
	\caption{Comparison of how well \omnijetalpha{} does on the generative task to the performance of a generative-only model, EPiC-FM-kintop. The latter does not use tokenization for the input data, and thus has access to the "real" input values. The main plots show the original \jetclass{} data, the reconstructed \jetclass{} data, as well as the generated events from \omnijetalpha{} and EPiC-FM-kintop. The ratio plots show how well the models perform with respect to \textit{their respective truths}.}
	\label{fig:gen_vs_bk}
\end{figure*}

To test the generative performance, the generative model was also trained on single-jet type data --- 10M jets each of \thad{} and \qcd{} --- separately. For these training, no tests of the task-transfer to classification were perfomed.

The result of the \qcd{} jet training is shown in \autoref{fig:gen_vs_truth_QCD}, of the \thad{} jet training in \autoref{fig:gen_vs_truth_top}. In the \qcd{} case, we see a good agreement between the reconstructed tokens and the generated events. However, it seems to be somewhat more difficult for the model to accurately model $\tau_{32}$ for \thad{} jets, which is also mirrored for this quantity in the combined model (see \autoref{fig:gen_vs_truth}). 

It is interesting to compare the result of \omnijetalpha{} with that of a different generative model. EPiC-FM~\cite{Birk:2023efj} was the first generative model trained on the \jetclass{} dataset, utilizing flow matching and operating without  tokenization. The result of the comparison can be seen in \autoref{fig:gen_vs_bk}. The plots show the \jetclass{} data, the reconstructed \jetclass{} token from this work, the EPiC-FM-kintop generated events, and the \omnijetalpha{} generated events. 
We use the more challenging \thad{} class for comparison.

The ratio plots under the main plots show the generated events compared to \textit{their respective truths}: direct \jetclass{} for EPiC-FM-kintop and Reconstructed \jetclass{} tokens for \omnijetalpha.
Hence, the ratios show how well the respective generative models learn to replicate their training data.
In general, we see that both models are doing well. \omnijetalpha{} has a somewhat higher discrepancy in the tails of all distributions except for constituent $\eta^{\mathrm{rel}}$ and the number of constituents. The most difficult distribution is the constituent $p_\mathrm{T}$, with bumps in the tail, which could also be seen in \autoref{fig:gen_vs_truth_top}. 

\vfill
\end{document}